\documentclass[prd,12pt,nofootinbib,superscriptaddress]{revtex4-2}
\usepackage{bm}
\usepackage{graphicx}
\usepackage{hyperref}
\usepackage{comment}
\usepackage{color}
\usepackage{orcidlink}
\usepackage{ulem}
\usepackage{subfigure}
\hypersetup{colorlinks=true,linkcolor=red,citecolor=blue}
\usepackage{xcolor}
\usepackage{amsmath,amsfonts,amssymb,amstext}
\usepackage{multirow}
\usepackage{float}
\usepackage{cases}
\usepackage{epsfig}%

\begin{document}
\title{Understanding curvature-matter interaction in viable $f(R)$ dark energy models: A dynamical analysis approach}
\author{Anirban Chatterjee \orcidlink{0000-0002-8904-7133}}
\email{Corresponding author: anirbanchatterjee@nbu.edu.cn \& iitkanirbanc@gmail.com}
\affiliation{Institute of Fundamental Physics and Quantum Technology, Department of Physics, School of Physical Science and Technology, Ningbo University, Ningbo, Zhejiang 315211, China}
\author{Yungui Gong \orcidlink{0000-0001-5065-2259}}
\email{gongyungui@nbu.edu.cn}
\affiliation{Institute of Fundamental Physics and Quantum Technology, Department of Physics, School of Physical Science and Technology, Ningbo University, Ningbo, Zhejiang 315211, China}

\begin{abstract}  
 We employ a linear stability analysis approach to explore the dynamics of matter and curvature-driven dark energy interactions within the framework of two types of viable $f(R)$ gravity models. The interaction is modeled via a source term in the continuity equations, $\mathcal{Q} =  \alpha \tilde{\rho}_{\rm m} \Big{(}\frac{3H^3}{\kappa^2 \rho_{\rm curv}} + \frac{\kappa^2 }{3H}\rho_{\rm curv} \Big{)}$. Our results reveal significant modifications to the fixed points and their stability criteria compared to traditional $f(R)$ gravity analyses without matter-curvature coupling. We identify constraints on model and coupling parameters necessary for critical point stability, illustrating how the interaction influences cosmic dynamics within specific parameter ranges. The findings are consistent with observed cosmic evolution, supporting stable late-time acceleration. Moreover, we highlight the coupling parameter's potential role in addressing the cosmic coincidence problem.
\end{abstract}

\maketitle

\section{Introduction}

Observations of Type Ia supernovae, specifically their redshifts and luminosity distances \cite{ref:Riess98, ref:Perlmutter}, provided compelling evidence that the universe transitioned from a phase of deceleration to one of accelerated expansion in its late-time evolution. This groundbreaking discovery has been corroborated by additional findings, including the temperature anisotropies observed in the cosmic microwave background data from the WMAP mission \cite{WMAP:2003elm,Hinshaw:2008kr}, and the detection of baryon acoustic oscillations \cite{SDSS:2005xqv}.  Within the framework of general relativity, this acceleration is attributed to the presence of an elusive component of energy density exhibiting negative pressure, commonly known as Dark Energy. Origin and nature of dark energy which are crucial to cosmic dynamics, remain a mystery in cosmology. In contrast, evidence for the existence of non-luminous matter, termed Dark Matter  (DM), comes from various astrophysical observations, including the rotation curves of spiral galaxies \cite{Sofue:2000jx}, gravitational lensing \cite{Bartelmann:1999yn}, the Bullet Cluster \cite{Clowe:2003tk}, and other colliding galaxy clusters. This matter interacts gravitationally, revealing its presence indirectly. Experiments such as WMAP \cite{Hinshaw:2008kr} and Planck \cite{Ade:2013zuv}  have shown that dark energy (DE) and dark matter (DM) together constitute about 96\% of the universe's total energy density today, with dark matter playing a significant role in the cosmic structure formation. \\

Einstein initially introduced the cosmological constant, $\Lambda g_{\mu\nu}$, to support a static universe, but discarded it after Hubble's expansion discovery. In the late 20th century, as cosmic acceleration was observed, it regained attention, leading to the $\Lambda$CDM model, where CDM stands for cold dark matter. However, this model faces the coincidence \cite{Zlatev:1998tr} and fine-tuning problems \cite{Martin:2012bt}, prompting the exploration of alternative dark energy models. Dark energy models modify Einstein's field equations by introducing a scalar field distinct from matter and radiation, such as quintessence \cite{Peccei:1987mm, Ford:1987de, Peebles:2002gy, Nishioka:1992sg, Ferreira:1997au, Ferreira:1997hj, Caldwell:1997ii, Carroll:1998zi, Copeland:1997et} and $k$-essence \cite{Fang:2014qga, ArmendarizPicon:1999rj, ArmendarizPicon:2000ah, ArmendarizPicon:2000dh, ArmendarizPicon:2005nz, Chiba:1999ka, ArkaniHamed:2003uy, Caldwell:1999ew, Bandyopadhyay:2017igc, Bandyopadhyay:2018zlz, Bandyopadhyay:2019ukl, Bandyopadhyay:2019vdd, Chatterjee:2022uyw}.  Other models, such as $f(R)$ gravity \cite{fr1,fr2,fr3,fr4,fr5,fr6,fr7,Chatterjee:2024rbh,Bertolami:2007gv,BarrosoVarela:2024ozs}, scalar-tensor theories, and braneworld models, modify the geometric structure of Einstein's equations to explain cosmic acceleration. These models often assume a homogeneous and isotropic universe described by the FLRW metric, though some consider inhomogeneous universes with perturbed FLRW metrics \cite{Chatterjee:2024duy}.\\

Dynamical systems methods are essential for investigating cosmic evolution in general cosmological models and analyzing specific cosmological solutions. Recent reviews have applied these techniques to evaluate the stability of various scalar field dark energy models \cite{Chatterjee:2021ijw, Chatterjee:2021hhj, Hussain:2022osn, Bhattacharya:2022wzu, Hussain:2023kwk}, as well as in the context of $f(R)$ gravity \cite{fr8, fr9, fr10, Bertolami:2009cd, Gunzig:2000ce, Odintsov:2017tbc, Lu:2019hra, Odintsov:2018uaw, Bahamonde:2017ize, Bahamonde:2019urw, Leon:2012mt, Xu:2012jf, Leon:2009dt, Leon:2009rc, Kofinas:2014aka, Basilakos:2019dof, Samart:2021viu, An:2015mvw}. A thorough exploration of dynamical systems in $f(R)$ gravity theories has been conducted in studies such as \cite{Samart:2021viu, Amendola:2006kh, Amendola:2006eh, Amendola:2007nt}, aiming to identify models that accurately represent cosmological evolution. These efforts have led to the development of cosmologically viable $f(R)$ gravity models, where the dynamics, governed by the Ricci scalar $R$, drive cosmic acceleration without the need for a cosmological constant or additional dark energy fields. These models also respect local gravity constraints and observational limits (as detailed in \cite{fr8, fr9, fr10}).\\

In this work, we explore the impact of curvature-driven dark energy interactions with dark matter on cosmological acceleration predictions within viable $f(R)$ gravity models, comparing them to non-interacting cases. $f(R)$ gravity theories with minimal matter coupling \cite{fr1,fr2,fr3,fr4,fr5} are governed by an action comprising a modified geometric term ($R \to f(R)$) and a usual matter term, with $g_{\mu\nu}$ ensuring minimal coupling to geometry. Modified field equation takes the standard form of  ${\cal G}_{\mu\nu} = 8\pi G T^{\rm (tot)}_{\mu\nu}$, where ${\cal G}_{\mu\nu}$ is the Einstein tensor and $T^{\rm (tot)}_{\mu\nu} = T^{\rm (M)}_{\mu\nu} + T^{\rm (curv)}_{\mu\nu}$. The curvature term $T^{\rm (curv)}_{\mu\nu}$, derived from $f(R)$ and its derivatives, vanishes for $f(R) = R$, restoring Einstein gravity. The matter term $T^{\rm (M)}_{\mu\nu}$ scales the usual energy-momentum tensor $\tilde{T}_{\mu\nu}$ by $F(R)^{-1}$ (where, $F(R)=\frac{df}{dR}$). $T^{\rm (curv)}_{\mu\nu}$ represents a stress-energy tensor for a curvature part, while $T^{\mu\nu}$ adjusts the matter and radiation tensor to incorporate $f(R)$ modifications. In the FLRW background, conservation of $T^{\rm (tot)}_{\mu\nu}$ leads to a continuity equation linking the energy density and pressure of the combined fluid (radiation, matter, and curvature). Source terms may appear in individual sector equations, provided the total fluid conservation remains intact. We assume radiation is decoupled from matter and curvature, with interactions between matter and curvature introducing a source term $\mathcal{Q}$ in their continuity equations, carrying opposite signs. $\mathcal{Q}$ represents the energy exchange rate between the two sectors. Prior work \cite{Samart:2021viu,  Chatterjee:2023oqo} defined $Q$ as proportional to the product of matter density and the Hubble parameter. Here, we refine it to $\mathcal{Q} =  \alpha \tilde{\rho}_{\rm m} \Big{(}\frac{3H^3}{\kappa^2 \rho_{\rm curv}} + \frac{\kappa^2 }{3H}\rho_{\rm curv} \Big{)}$. where $\alpha$ is the coupling strength, and $\tilde{\rho}_{\rm m}$ and $\rho_{\rm curv}$ are the energy densities of matter and curvature, respectively.\\

Curvature-matter interactions introduce the coupling parameter $\alpha$, which, along with $f(R)$ parameters, significantly impacts the evolution equations and dynamics in the coupled scenario. Using dynamical analysis, we investigated matter-curvature interactions in viable $f(R)$ models by deriving autonomous equations, identifying fixed points, and analyzing their stability, crucial for understanding cosmic evolution.
We studied two $f(R)$ models: the generalized $\Lambda$CDM model, $f(R) = (R^b - \Lambda)^c$ with $c \geqslant 1$ and $bc \approx 1$, and the power-law model, $f(R) = R - \gamma R^n$ with $\gamma > 0$ and $0 < n < 1$, both capable of driving cosmic acceleration while remaining viable. Stability of fixed points depends on parameters like  $(b, c)$, $n$ and $\alpha$. Choosing these parameters values, we found that the critical point's nature switched from saddle to stable and vice versa, highlighting the importance of analyzing the parameter space of each model with interactions. We investigate the impact of curvature-driven dark energy on different phases of cosmic evolution. This includes examining the evolution of the matter-to-curvature energy density ratio ($r_{mc}$), deceleration parameter ($q$)  and jerk parameter ($j$). Evolution of $r_{mc}$ parameter reveals how curvature and matter dominance shift throughout cosmic history, while $q$ and $j$ offer detailed insights into the kinematics of the evolution. Our findings suggest that all models eventually converge to the $\Lambda$CDM model in the far future, highlighting the scenarios possible within the context of matter-curvature interactions.\\

The paper is structured as follows: In Sec.\ [\ref{Sec:2}], we present the theoretical framework for the interaction between curvature and matter in a flat FLRW metric. Sec.\ [\ref{Sec:3}] formulates the autonomous equations of the dynamical system, incorporating matter-curvature-driven dark energy interactions in $f(R)$ gravity, and discusses the two $f(R)$ models considered. We analyze the fixed points and their stability for each model and explore the impact of curvature-driven dark energy on cosmic evolution, along with the effect of the coupling parameter in addressing the cosmic coincidence problem in both models. In Sec.\ [\ref{Sec:4}] we present the conclusions of this study.

\section{Framework for analyzing curvature-matter interactions in $ f(R) $ gravity}
\label{Sec:2}

Action corresponding to minimal curvature-matter coupling in the context of $f(R)$ gravity can be written as,
\begin{eqnarray}
S = \frac{1}{2\kappa^2}\int d^4x\sqrt{-g}f(R) + \int d^4x~L_M(g_{\mu \nu},\phi _M)\,,
\label{eq:b1}
\end{eqnarray}
In this expression, $\kappa^2 = 8\pi G$, $f(R)$ represents a general function of the Ricci scalar $R$, and $L_{M}$, referred to as the matter Lagrangian, describes the Lagrangian density for the universe's radiation and matter components (including baryonic matter and cold dark matter (CDM)). The determinant of the space-time metric tensor $g_{\mu\nu}$ is denoted by $g$. In metric formalism, varying this action with respect to the metric tensor $g_{\mu\nu}$ yields the corresponding modified field equations as follows:

\begin{eqnarray}
F(R) R_{\mu\nu}-\frac{1}{2}g_{\mu \nu}f(R) + g_{\mu \nu}\square F(R) - \nabla_{\mu}\nabla_{\nu} F(R)  = \kappa^2 \tilde{T}_{\mu\nu}^{(\rm M)}\,.
\label{eq:b2}
\end{eqnarray}
In this context, $F(R) \equiv \frac{df}{dR}$, and $\tilde{T}_{\mu \nu}^{(\rm M)}$ represents the stress-energy tensor for the radiation and matter components, expressed as:
\begin{eqnarray}
\tilde{T}_{\mu \nu}^{(\rm M)} \equiv -{2 \over \sqrt{-g}}{\delta(\sqrt{-g}L_M)\over \delta(g^{\mu\nu})} \,.
\label{eq:b3}
\end{eqnarray}
Eq. \eqref{eq:b2} can be rewritten as 
\begin{eqnarray}
 {\cal G}_{\mu\nu} \equiv R_{\mu\nu}- \frac{1}{2}Rg_{\mu \nu} 
 = \kappa^2 \left(T_{\mu\nu}^{(\rm M)} + T_{\mu\nu}^{(\rm curv)}\right) 
 \equiv \kappa^2 T_{\mu\nu}^{(\rm tot)} \,,
\label{eq:b4}
\end{eqnarray}
where, ${\cal G}_{\mu\nu}$ denotes Einstein tensor and
\begin{eqnarray}
 T_{\mu\nu}^{(\rm curv)} & \equiv & 
\frac{1}{\kappa^2 F} 
\left[\frac{1}{2}(f-RF)g_{\mu \nu} +  \left(\nabla_{\mu}\nabla_{\nu}- g_{\mu \nu}\square \right)F 
\right] 
\label{eq:b5} \\
\mbox{and} \quad T_{\mu\nu}^{(\rm M)}
& \equiv &
\frac{1}{F} \tilde{T}_{\mu\nu}^{(\rm M)}\label{eq:b6} 
\end{eqnarray}

In the revised form, the modified field eqn.  \eqref{eq:b4} represents a  FLRW universe
 containing a fluid, with the complete energy-momentum tensor given as
 $T_{\mu\nu}^{(\rm tot)} \equiv T_{\mu\nu}^{(\rm M)} + T_{\mu\nu}^{(\rm curv)}$. 
  The influence of the function $f(R)$ is apparent in both components $T_{\mu\nu}^{(\rm curv)}$ and $T_{\mu\nu}^{(\rm M)}$, which together constitute $T_{\mu\nu}^{(\rm tot)}$. The component $T_{\mu\nu}^{(\rm curv)}$ depends exclusively on $f(R)$ and its higher derivatives, becoming zero when $f(R) = R$. This term effectively acts as the stress-energy tensor, representing a fluid-like description of the curvature function $f(R)$. On the other hand, the component $T_{\mu\nu}^{(\rm M)}$ is obtained by rescaling $\tilde{T}_{\mu\nu}^{(\rm M)}$ with a functional multiplier $\frac{1}{F}$, which reduces to unity when $f(R) = R$. This represents the modified stress-energy tensor for matter and radiation, accounting for gravitational modifications introduced by $f(R)$. In a flat FLRW spacetime background, modeling matter and radiation as a perfect fluid leads to the (unmodified) stress-energy tensor $\tilde{T}_{\mu\nu}^{(\rm M)}$ being diagonal, with its components determined entirely by the combined energy densities $(\tilde{\rho}_{\rm m} + \tilde{\rho}_{\rm r})$ of dark matter ($\tilde{\rho}_{\rm m}$) and radiation ($\tilde{\rho}_{\rm r}$), along with the radiation pressure $\tilde{P}_{\rm r} = \tilde{\rho}_{\rm r}/3$. The matter component is treated as non-relativistic, pressureless dust. Consequently, the modified stress-energy tensor $T_{\mu\nu}^{(\rm M)} = \frac{1}{F} \tilde{T}_{\mu\nu}^{(\rm M)}$ corresponds to a fluid with energy density $(\rho_{\rm m} + \rho_{\rm r})$ and pressure $P_{\rm r}$, where 
$(\rho_{\rm r}, \rho_{\rm m}, P_{\rm r}) \equiv \left(\frac{\tilde{\rho}_{\rm r}}{F}, \frac{\tilde{\rho}_{\rm m}}{F}, \frac{\tilde{P}_{\rm r}}{F}\right)$. It is crucial to highlight that $\rho_i$ and $P_{\rm r}$ (for $i = \rm{r}, \rm{m}$) remain positive due to the inherent positivity of $(\tilde{\rho}_i, \tilde{P}_{\rm r})$ and the requirement that $F > 0$, which is a necessary condition for any cosmologically viable $f(R)$ model. Based on these considerations, the modified field equation \eqref{eq:b4} leads to the following modified versions of the Friedmann equations for the `00' and `$ii$' components, respectively: 

\begin{eqnarray}
 H^2 
= \frac{\kappa^2 }{3}\Big{[}  \rho_{\rm r}  + \rho_{\rm m} +   \rho_{\rm curv}  \Big{]}\,
\label{eq:b7}\\
 \dot{H} +  H^2 = - \frac{\kappa^2}{6}
\Big{[}    \rho_{\rm r}  + \rho_{\rm m}  + \rho_{\rm curv}  +   3(P_{\rm r}  + P_{\rm curv})\Big{]} 
\label{eq:b8}\,
\end{eqnarray}
Hubble parameter $(H) \equiv \dot{a}/a$ and $a$ is the scale factor of FLRW metric. Expression for curvature energy density ($\rho_{\rm curv}$) and pressure ($P_{\rm curv}$) are,
 \begin{eqnarray}
 \rho_{\rm curv} & \equiv &  
 \frac{1}{\kappa^2 F}\left(\frac{RF-f}{2}-3H\dot{R}F'\right)
\label{eq:b9}\\
 P_{\rm curv}& \equiv &\frac{1}{\kappa^2 F}\left(\dot{R}^2F'' + 2H\dot{R}F'+\ddot{R}F' +\frac{1}{2}(f-RF)\right)
\label{eq:b10}\,
\end{eqnarray}
Here, $'$ and $''$ correspond to the higher derivative in respect to $R$. In a flat FLRW spacetime, eqn. \eqref{eq:b5} implies that the stress-energy tensor $T_{\mu\nu}^{\rm curv}$, associated with the curvature fluid, behaves like that of an ideal fluid. The energy density and pressure, $\rho_{\rm curv}$ and $P_{\rm curv}$, are defined in eqs. \eqref{eq:b9} and \eqref{eq:b10}, respectively. Additionally, eqn. \eqref{eq:b7} can be reformulated as:
\begin{eqnarray}
 \Omega_{\rm r} + \Omega_{\rm m} + \Omega_{\rm curv} &=& 1
\label{eq:b11}
\end{eqnarray}

Parameters  $\Omega_{r}$, $\Omega_{\rm m}$ and $\Omega_{\rm curv}$ denote the modified density parameters, expressed as follows:
\begin{eqnarray}
\Omega_{\rm r} \equiv \frac{\kappa^2 \rho_{\rm r}}{3H^2} = \frac{\kappa^2\tilde{\rho}_{\rm r}}{3FH^2}\,,\quad
\Omega_{\rm m} \equiv \frac{\kappa^2\rho_{\rm m}}{3H^2} = \frac{\kappa^2\tilde{\rho}_{\rm m}}{3FH^2}\,,\quad
 \Omega_{\rm curv} \equiv \frac{\kappa^2\rho_{\rm curv}}{3H^2}
\label{eq:b12}
\end{eqnarray}

It is important to note that for the special case where $f(R) = R$ ($F = 1$, $F' = F'' = 0$), we have $\rho_{\rm curv} = 0$ and $P_{\rm curv} = 0$. In this scenario, the energy density and pressure of radiation are given by $(\rho_i, P_{\rm r}) = (\tilde{\rho}_i, \tilde{P}_{\rm r})$. This leads to the modified Friedmann eqs. \eqref{eq:b7} and \eqref{eq:b8} reducing to the standard Friedmann equations.\\

 By taking the divergence of both sides of equation \eqref{eq:b4} and applying Bianchi's identity, we obtain the conservation equation for the total stress-energy tensor. Conservation equation for the total stress-energy tensor is derived by taking the divergence of both sides of eqn. \eqref{eq:b4} and applying Bianchi's identity with $T_{\mu\nu}^{\rm tot} = T^{(M)}_{\mu\nu} + T^{\rm curv}_{\mu\nu}$:
\begin{eqnarray}
\nabla^{\mu}  T_{\mu\nu}^{\rm tot} &=& \nabla^{\mu}  \left(T_{\mu\nu}^{(\rm M)} + T_{\mu\nu}^{(\rm curv)}\right) =0 
\label{eq:b13}
\end{eqnarray}

In the FLRW spacetime, eqn. \eqref{eq:b13} assumes the form of a continuity equation for the total fluid, where the total energy density is defined as ${\rho}_{\rm tot} \equiv \rho_{\rm r} + \rho_{\rm m} + \rho_{\rm curv}$ and the total pressure as $P_{\rm tot} \equiv P_{\rm r} + P_{\rm curv}$,

\begin{eqnarray}
\dot{\rho}_{\rm tot}+3H{\rho}_{\rm tot}(1 + \omega_{\rm tot}) = 0\,
\label{eq:b14}
\end{eqnarray}

In this context, the total equation of state (EoS) parameter, defined as $\omega_{\rm tot} \equiv P_{\rm tot}/\rho_{\rm tot}$, characterizes the composite sectors, which include radiation, matter, and curvature. Although the total stress-energy tensor $T^{(\rm tot)}_{\mu\nu}$ is conserved, this conservation does not hold for the individual components $T^{(\rm curv)}_{\mu\nu}$ and $T^{(M)}_{\mu\nu}$. The latter further divides into the sub-components for radiation, $T^{(\rm r)}_{\mu\nu}$, and matter, $T^{(\rm m)}_{\mu\nu}$. This lack of separate conservation enables interactions between the matter and curvature sectors.

Considering that the radiation component has a negligible effect during the later stages of cosmic evolution, we assume no interactions between radiation and the matter-curvature sectors, while allowing for interactions between the matter and curvature sectors. Based on these assumptions, we interpret the conservation eqn. \eqref{eq:b13} as a system of equations:
   (a) $  \nabla^{\mu} T_{\mu\nu}^{(\rm r)} = 0 $, 
   (b) $ \nabla^{\mu} T_{\mu\nu}^{(\rm m)} = - \nabla^{\mu} T_{\mu\nu}^{(\rm curv)} \equiv - Q_{\nu} \neq 0 $,
These equations together ensure the total stress-energy tensor \( T^{(\rm tot)}_{\mu\nu} \) is conserved. In FLRW spacetime, equation (a) implies \( \dot{\rho}_{\rm r} + 4H \rho_{\rm r} = 0 \) and \( \rho_{\rm r} \sim a^{-4} \), while equation set (b) corresponds to non-conservation equations of the form:
\begin{eqnarray}
\dot\rho_{\rm m} +3H\rho_{\rm m} &=& - \mathcal{Q}\label{eq:b15}\\
\dot{\rho}_{\rm curv}+3H\left(\rho_{\rm curv}+P_{\rm curv}\right)&=& \mathcal{Q} \label{eq:b16}
\end{eqnarray}
Based on these considerations, we propose formulating the conservation equation as a system of equations with a source term \( \mathcal{Q} \), which represents a time-dependent function quantifying the rate of energy exchange between the curvature and matter sectors, thus encapsulating their interactions.

The interaction between dark matter and curvature-driven dark energy can be introduced either at the Lagrangian level, guided by field-theoretic principles, or at the phenomenological level by defining the non-zero source term $\mathcal{Q}$ (in eqs. \eqref{eq:b15} and \eqref{eq:b16}) in terms of appropriate cosmological quantities. In this work, we adopt the phenomenological approach and specify $\mathcal{Q}$ as:  
$\mathcal{Q} =  \alpha \tilde{\rho}_{\rm m} \Big{(}\frac{3H^3}{\kappa^2 \rho_{\rm curv}} + \frac{\kappa^2 }{3H}\rho_{\rm curv} \Big{)}$. Dimensionless coupling parameter $\alpha$ is introduced to maintain dimensional consistency in both equations.  In this context, it is worth mentioning that a previous study \cite{Chatterjee:2023oqo, Samart:2021viu} investigated the impact of interaction using a source term that depended solely on a combination of the energy densities of the matter and curvature sectors, the Hubble parameter, and the coupling constant. Building on this framework, our selected form of the source term exhibits distinct scaling behavior across cosmic epochs, making it useful for explaining the radiation-to-dark energy transition. During the radiation-dominated era, the term $( H^3 / \rho_{\rm curv} )$ dominates and scales as $(1+z)^6$, ensuring minimal interaction influence on radiation while allowing energy exchange between matter and curvature. In the matter-dominated era, the interaction facilitates a gradual transfer of energy from matter to curvature. As the universe enters the dark energy-dominated phase, the second term $( H \rho_{\rm curv})$ governs, sustaining curvature dominance and driving cosmic acceleration. This scaling behavior ensures the coupling term adapts dynamically across different epochs, transferring energy between matter and curvature, and providing a unified framework to model cosmic evolution.  

We impose $\rho_{\rm curv} > 0$ to interpret it as curvature’s fluid-equivalent energy density, restricting $f(R)$ models to $RF - f > 6H\dot{R}F'$ as shown in eqn. \eqref{eq:b9}. This constrains $\alpha$, as $\rho_{\rm curv}$ depends on $\mathcal{Q}$ as mentioned in eqn. \eqref{eq:b16}. The modified density parameters $\Omega_{\rm r}$, $\Omega_{\rm m}$, and $\Omega_{\rm curv}$ remain positive and satisfy eqn. \eqref{eq:b11}. Varying $0 \leq \Omega_{\rm m} \leq 1$ explores curvature-matter interactions, imposing constraints on model parameters. Accelerating phase ($\ddot{a} > 0$) requires $\omega_{\rm tot} < -1/3$ as given in  eqn. \eqref{eq:b8}. $\omega_{\rm tot}$, influenced by $F(R)$ and its derivatives via $\rho_{\rm curv}$ and $P_{\rm curv}$ as given in eqns. \eqref{eq:b15}, \eqref{eq:b16}, allows $-1 < \omega_{\rm tot} < -1/3$ for non-phantom dark energy, with $\alpha$ and $F(R)$ parameters consistent with late-time dark energy dynamics.  A comprehensive analysis of solar system tests\ [\ref{Sec:Apna}] and the conformal transformation in the context of $f(R)$ gravity models\ [\ref{Sec:Apnb}] are provided in the appendices.

\section{Dynamical analysis of $f(R)$ models with curvature-matter couplings}
\label{Sec:3}
\subsection{Construction of Autonomous equations}
\label{Sec:3.1}

 We use dynamical analysis to investigate curvature-matter interactions in cosmologically viable $f(R)$ gravity models. In this subsection, we present the set of autonomous equations formulated in terms of basic dynamical variables, which define the cosmological evolution in this framework, laying the foundation for the subsequent dynamical analysis. \\

Dynamical variables can be defined as, 
\begin{eqnarray}
X_{1}= - \frac{\dot F}{HF}, \quad  X_{2}= - \frac{f}{6FH^2}, \quad X_{3}=\frac{R}{6H^2},  \quad X_{4} =\frac{\kappa^2  \rho_{\rm r} }{3H^2} = \frac{\kappa^2  \tilde{\rho}_{r}  }{3FH^2}=\Omega_{r}
\label{eq:d0}
\end{eqnarray}

To capture temporal variations, we introduce the dimensionless parameter $N = \ln a$ and utilize the defined dynamical variables ($X_i$'s). With this approach, Eqs. \eqref{eq:b9}, \eqref{eq:b15}, and \eqref{eq:b16}, which describe the evolution of the system in the context of curvature-matter interactions, can be rewritten as a set of four autonomous equations, forming a 4-dimensional dynamical system.

\begin{eqnarray}
\frac{dX_{1}}{dN}&=& -1 - X_{1}X_{3}- X_{3} - 3X_{2} + X_{4} + X_{1}^2 \nonumber\\
&&\qquad +\ \alpha \left(1- X_{1} - X_{2} - X_{3} - X_{4}\right)
 \left( \frac{1}{(X_{1} + X_{2} + X_{3})}+(X_{1} + X_{2} + X_{3}) \right) \label{eq:d1} \\
\frac{dX_{2}}{dN} &=& \frac{X_{1}X_{3}}{m}-X_{2}\left(2X_{3}-4-X_{1}\right)  \label{eq:d2}\\
\frac{dX_{3}}{dN} &=& -\frac{X_{1}X_{3}}{m} - 2X_{3}\left( X_{3}-2\right) \label{eq:d3}\\
\frac{dX_{4}}{dN} &=& - 2X_{3}X_{4} + X_{1}X_{4} \,, \label{eq:d4}
\end{eqnarray}
where,
\begin{eqnarray}
m &\equiv & \frac{d \ln F}{d \ln R}= \frac{R F' }{F}\,,
\label{eq:d5}
\end{eqnarray}
Additionally, a new parameter, $r$, is defined as,
\begin{eqnarray}
 r & \equiv & -\frac{d \ln f}{d \ln R}= -\frac{R F}{f} = \frac{X_3}{X_2}\,.
\label{eq:d6}
\end{eqnarray}
Both $m$ and $r$ are dimensionless parameters that depend on $R$ through $f(R)$, allowing $m$ to be expressed as $m = m(r)$. Each distinct form of this relationship $m = m(r)$ corresponds to a unique class of $f(R)$ models. Setting the coupling parameter $\alpha$ to zero reduces the system to autonomous equations that exclude curvature-matter interactions. Moreover, for $f(R) = R$, the parameter $m$ (and thus $X_1$) tends to zero, invalidating the construction of the dynamical system. Using eqs.~\eqref{eq:b11} and \eqref{eq:b12}, the dynamical variables in Eq.~\eqref{eq:d0} are subject to the constraint given by the equation.
\begin{eqnarray}
\Omega_{\rm m}   = 1 - X_{1} - X_{2} - X_{3} - X_{4} 
\label{eq:d7}
\end{eqnarray}
%
where $\Omega_{\rm m}$ satisfies the inequality $0 \leqslant \Omega_{\rm m} \leqslant 1$, and we can further express,
\begin{eqnarray}
\Omega_{\rm curv}  =  1 - \Omega_{\rm m} -  \Omega_{r} =  X_{1} + X_{2} + X_{3}\,.
\label{eq:d8}
\end{eqnarray}
Additionally, total equation of state (EoS) parameter $\omega_{\rm tot}$ can be written as, 
\begin{eqnarray}
\omega_{\rm tot} &=& -1 - \frac{2\dot{H}}{3H^2} = -\frac{1}{3}(2X_{3}-1)
\label{eq:d9}
\end{eqnarray}
%
Ratio of matter to curvature energy density, $r_{mc} \equiv \Omega_{\rm m}/\Omega_{\rm curv}$, the deceleration parameter, $q \equiv - a\ddot{a}/\dot{a}^2$ and jerk parameter $j \equiv \frac{\dddot{a}}{aH^3}$, play an important role in evaluating cosmic dynamics. In the later stages of the universe, $r_{mc}$ roughly corresponds to the ratio of dark matter to dark energy densities, with observations indicating that dark matter's contribution far outweighs that of baryonic matter in the present-day universe. As such, $r_{mc}$ serves as a potential solution to the cosmic coincidence problem, which pertains to the near equality of dark matter and dark energy densities. Deceleration parameter $q$ reflects the kinematic characteristics of cosmic expansion. A negative value of $q$ indicates the acceleration of the universe, a hallmark of dark energy's dominance, and when $q = -1$, it corresponds to the cosmological constant. The transition from a positive to a negative $q$ value marks the shift from deceleration to acceleration in cosmic expansion. Fluctuations in $q$ are indicative of dynamical dark energy models, which capture the evolving acceleration of the universe. Jerk parameter, $j$, represents the variation in acceleration over time (i.e., the derivative of $q$), offering a more detailed perspective on the evolution of cosmic acceleration. Therefore, these parameters are vital in understanding dark energy’s role across different stages of cosmic history and in distinguishing between various models that explain cosmic acceleration. Within the framework of the matter-curvature interaction scenario explored in this study, the parameters $r_{mc}$, $q$  and $j$ can be expressed in terms of the dynamical variables ($X_i$'s) as:
\begin{eqnarray}
 r_{mc} &\equiv & \frac{\Omega_{\rm m}}{\Omega_{\rm curv}} 
 = \frac{1 - X_1 - X_2 - X_3 - X_4}{X_1 + X_2 + X_3} \label{eq:d10}\\
q &\equiv & - \frac{a\ddot{a}}{\dot{a}^2} = -1 - \frac{\dot{H}}{H^2} = 1 - X_3  \label{eq:d11}  \\
 j &\equiv & \frac{\dddot{a}}{aH^3} = -\frac{X_1 X_3}{m} + 2(1-X_3)^2 + (1-X_3)-2X_3(X_3-2) \label{eq:d12} 
\end{eqnarray}

 The fixed points of the 4D dynamical system are defined by $dX_i/dN = 0$ ($i = 1, \ldots, 4$) and are critical for understanding cosmic evolution driven by curvature-matter interactions. Linear stability analysis, applied near these points, involves a first-order Taylor expansion of the autonomous equations $d\vec{X}/dN = \vec{f}(\vec{X})$, where $\vec{X} = \{X_1, X_2, X_3, X_4\}$ and $\vec{f} = \{f_1, f_2, f_3, f_4\}$, derived from eqs.~\eqref{eq:d1}, \eqref{eq:d2}, \eqref{eq:d3}, \eqref{eq:d4}. The Jacobian matrix $J = \partial \vec{f} / \partial \vec{X}$, a $4 \times 4$ matrix, determines the stability of the fixed points via its eigenvalues. Examining the eigenvalues of the Jacobian matrix $J$ at the critical points determines the stability of the fixed points. Fixed points are asymptotically stable (unstable) if all eigenvalues have negative (positive) real parts, while mixed signs in the real parts of any eigenvalue pair classify the point as a saddle. If an eigenvalue approaches zero, linear stability theory becomes insufficient, requiring center manifold theory for further analysis. In this study, however, no non-hyperbolic fixed points were identified across the $f(R)$ models analyzed. Consequently, linear stability analysis was sufficient for characterizing the fixed points without needing advanced methods.

\subsection{Selection of modified $f(R)$ gravity models}
\label{Sec:3.2}

The dynamical analysis presented here is constructed around $f(R)$-driven dark energy models, emphasizing the interaction between the curvature and matter sectors. We specifically select models that not only induce cosmic acceleration but also adhere to cosmological viability. Under the metric formalism, any viable $f(R)$ function must fulfill stringent conditions, as detailed in the references \cite{fr8,fr9,fr10,Faraoni:2008mf,Tsujikawa:2010sc,Li:2007xn}.  Our study examines the autonomous system by focusing on two distinct cases. The first case assumes an $f(R)$ function that produces a constant $m$, while the second involves an $f(R)$ function where $m$ varies as a function of $r$, denoted by $m(r)$. These two cases are referred to as (I) and (II), and their specifics are discussed below.

%
%

\begin{itemize}
\item[(I)]   
A constant $m$ scenario arises from the modified gravity model $f(R) = (R^b - \Lambda)^c$ with $(c \geqslant 1, bc \approx 1)$. This generalized $\Lambda$CDM model satisfies local gravity constraints and converges to standard $\Lambda$CDM with $m=0$ for $c \geqslant 1$ and $bc \approx 1$. The model's parameters are:
$m = \frac{(bc - 1)R^b - b\Lambda + \Lambda}{R^b - \Lambda}, \quad r = -\frac{bcR^b}{R^b - \Lambda}$ and $m(r) = \left(\frac{1 - c}{c}\right)r + b - 1$. A dynamical analysis shows that at stable points, $r = -1 - m$, leading to $m = -1 + bc$, a constant.

\item[(II)] This scenario considers $m$ as a function of $r$, given by $m(r) = \frac{n(1+r)}{r}$, which is realized for the power-law form $f(R) = R - \gamma R^n$ with $(\gamma > 0, 0 < n < 1)$. We refer to this as the `power law model'. This model correctly describes cosmic evolution in non-interacting curvature-matter scenarios and satisfies the conditions for cosmological viability \cite{Li:2007xn} in the specified range of $\gamma$ and $n$. For this $f(R)$, we have $m = \frac{\gamma(n-1)nR^n}{R - \gamma n R^n}$ and $r = \frac{R - \gamma n R^n}{R - \gamma R^n}$, leading to $m(r) = \frac{n(1+r)}{r}$ after eliminating $\gamma$.

\end{itemize}

The selection of the two cases in this study: (I) $m = -1 + bc$ (constant) and (II) $m = m(r) = \frac{n(1+r)}{r}$, aims to examine the effects of the two $f(R)$-models in the presence of matter-curvature interactions. We analyze the stability of various modified gravity models within the framework of a coupled curvature-matter system using linear stability theory.

\subsection{Fixed point stability analysis in the Generalized $\Lambda$CDM Model}

\begin{table}[H]
\centering
\begin{tabular}{|c|c|c|c|}
\hline
 Critical  & \multirow{2}{*}{($X_1,X_2,X_3,X_4$)} & \multirow{2}{*}{$\Omega_{\rm m} $} & \multirow{2}{*}{$\omega_{\rm tot}$} \\
points & & &  \\
\hline\hline
$P_{1}$ & ($-4$ , 5 , 0 , 0) & 0 & $\frac13$  \\
\hline
$P_{2}$ & (0 , $-1$ , 2 , 0) & 0 & $-1$  \\
\hline
\multirow{2}{*}{$P_{3 \pm}$} & \multirow{2}{*}{$\left( -4 , \frac{-3 +8\alpha \mp \sqrt{9-4\alpha^2} }{2\alpha} , 0 , 0 \right)$} & \multirow{2}{*}{$5-\frac{-3\mp8 \alpha  \sqrt{9-4 \alpha ^2}}{2 \alpha }$} & \multirow{2}{*}{$\frac13$}  \\
&&&\\
\hline
\multirow{2}{*}{$P_{4 \pm}$} & \multirow{2}{*}{$\left(0 , \frac{- 3 - 4\alpha \mp \sqrt{9-4\alpha^2}}{2\alpha} , 2 , 0\right)$} & \multirow{2}{*}{$-1-\frac{-3\mp 4 \alpha  \sqrt{9-4 \alpha ^2}}{2 \alpha }$} & \multirow{2}{*}{$-1$} \\
&&&\\
\hline
\multirow{2}{*}{$P_{5 \pm}$} & \multirow{2}{*}{$\left(\frac{1 \mp \sqrt{-4 \alpha ^2+4 \alpha +1}}{2 (\alpha -1)} , 0 , 0 , 0\right)$} & \multirow{2}{*}{$ 1 -  \frac{1 \mp \sqrt{-4 \alpha ^2+4 \alpha +1}}{2 (\alpha -1)}$} & \multirow{2}{*}{$\frac13$} \\
&&&\\
\hline
\multirow{2}{*}{$P_6$} & \multirow{2}{*}{$\left (1, 0 , 0 , 0\right)$} & \multirow{2}{*}{$ 0$} & \multirow{2}{*}{$\frac13$} \\
&&&\\
\hline
\multirow{2}{*}{$P_7$} & \multirow{2}{*}{$\left(-\frac{2 (\text{bc}-2)}{2 \text{bc}-1}, \frac{5-4 \text{bc}}{2 \text{bc}^2-3 \text{bc}+1} , \frac{4 \text{bc}^2-5 \text{bc}}{2 \text{bc}^2-3 \text{bc}+1} , 0\right)$} & \multirow{2}{*}{$ 0$} & \multirow{2}{*}{$\frac{-6 \text{bc}^2+7 \text{bc}+1}{6 \text{bc}^2-9 \text{bc}+3}$} \\
&&&\\
\hline
\multirow{2}{*}{$P_8$} & \multirow{2}{*}{$P_8(X_1, X_2, X_3, 0$)} & \multirow{2}{*}{$ P_8 (\Omega_m) $} & \multirow{2}{*}{$P_8(\omega_{\rm tot.})$} \\
&&&\\
\hline
\multirow{2}{*}{$P_9$} & \multirow{2}{*}{$P_9(X_1, X_2, X_3, 0$)} & \multirow{2}{*}{$ P_9 (\Omega_m) $} & \multirow{2}{*}{$P_9(\omega_{\rm tot.})$} \\
&&&\\
\hline

\multirow{2}{*}{$P_{10}$} & \multirow{2}{*}{$\left(\frac{4 (\text{bc}-1)}{\text{bc}}, -\frac{2 (\text{bc}-1)}{\text{bc}^2}, \frac{2 (\text{bc}-1)}{\text{bc}}, \frac{-5 \text{bc}^2+8 \text{bc}-2}{\text{bc}^2}\right)$} & \multirow{2}{*}{$ 0 $} & \multirow{2}{*}{$\frac{4}{3 \text{bc}}-1$} \\
&&&\\
\hline

\end{tabular}
\caption{Critical points of generalized $\Lambda$CDM model with corresponding $\Omega_{\rm m}$ and $\omega_{\rm tot}$. Details of the critical points $P_8$ and $P_9$ are provided below.} 
 
\label{tab:1}
\end{table}

{\scriptsize   $P_8(X_1, X_2, X_3, 0$) =\\  \Big{(}$-\frac{-4 \alpha +(8 \alpha +2) \text{bc}^3-5 (4 \alpha +1) \text{bc}^2+\sqrt{-\left((\text{bc}-1) \text{bc}^2 \left(-4 \alpha ^2+4 \left(4 \alpha ^2-4 \alpha -25\right) \text{bc}^3+8 \left(-4 \alpha ^2+\alpha +20\right) \text{bc}^2+\left(20 \alpha ^2-69\right) \text{bc}+9\right)\right)}+(16 \alpha +3) \text{bc}}{(2 \text{bc}-1) \left(\alpha +2 (\alpha -1) \text{bc}^2-3 \alpha  \text{bc}\right)},\\
 \frac{-16 (\alpha -1) \text{bc}^4+(40 \alpha -26) \text{bc}^3+(13-32 \alpha ) \text{bc}^2-\sqrt{-\left((\text{bc}-1) \text{bc}^2 \left(-4 \alpha ^2+4 \left(4 \alpha ^2-4 \alpha -25\right) \text{bc}^3+8 \left(-4 \alpha ^2+\alpha +20\right) \text{bc}^2+\left(20 \alpha ^2-69\right) \text{bc}+9\right)\right)}+(8 \alpha -3) \text{bc}}{2 (\text{bc}-1) \text{bc} (2 \text{bc}-1) \left(\alpha +2 (\alpha -1) \text{bc}^2-3 \alpha  \text{bc}\right)},\\
 \frac{16 (\alpha -1) \text{bc}^4+(26-40 \alpha ) \text{bc}^3+(32 \alpha -13) \text{bc}^2+\sqrt{-\left((\text{bc}-1) \text{bc}^2 \left(-4 \alpha ^2+4 \left(4 \alpha ^2-4 \alpha -25\right) \text{bc}^3+8 \left(-4 \alpha ^2+\alpha +20\right) \text{bc}^2+\left(20 \alpha ^2-69\right) \text{bc}+9\right)\right)}+(3-8 \alpha ) \text{bc}}{2 (\text{bc}-1) (2 \text{bc}-1) \left(\alpha +2 (\alpha -1) \text{bc}^2-3 \alpha  \text{bc}\right)}, 
 0 \Big{)}$,\\
  \scriptsize  $P_8(\Omega_m)$= 
  $
\frac{
\sqrt{\text{bc}^2 \left(-4 \alpha^2 \left(2 \text{bc}^2 - 3 \text{bc} + 1\right)^2 
+ 8 \alpha (\text{bc} - 1) (2 \text{bc} - 1) \text{bc}^2 
+ \left(10 \text{bc}^2 - 13 \text{bc} + 3\right)^2 \right)} 
\\
+ \text{bc} \left(2 \alpha + \text{bc} \left(-6 \alpha + (4 \alpha + 6) \text{bc} - 13\right) + 3 \right)
}
{2 \text{bc} \left(\alpha + 2 (\alpha - 1) \text{bc}^2 - 3 \alpha \text{bc}\right)}
$,\\
   \scriptsize  $P_8(\omega_{\rm tot. })$=$\frac{
\alpha - \sqrt{\text{bc}^2 \left(-4 \alpha^2 \left(2 \text{bc}^2 - 3 \text{bc} + 1\right)^2 
+ 8 \alpha (\text{bc} - 1) (2 \text{bc} - 1) \text{bc}^2 
+ \left(10 \text{bc}^2 - 13 \text{bc} + 3\right)^2 \right)} 
+ \text{bc} \left(2 \alpha + \text{bc} \left(-19 \alpha - 4 \text{bc} \left(-7 \alpha + 3 (\alpha - 1) \text{bc} + 5\right) + 11 \right) - 3\right)
}{
3 (\text{bc} - 1) (2 \text{bc} - 1) \left(\alpha + 2 (\alpha - 1) \text{bc}^2 - 3 \alpha \text{bc}\right)
}$.\\
\vspace{1cm}

\scriptsize $P_9(X_1, X_2, X_3, 0$) =\\  \Big{(} $\frac{4 \alpha -2 (4 \alpha +1) \text{bc}^3+5 (4 \alpha +1) \text{bc}^2+\sqrt{-\left((\text{bc}-1) \text{bc}^2 \left(-4 \alpha ^2+4 \left(4 \alpha ^2-4 \alpha -25\right) \text{bc}^3+8 \left(-4 \alpha ^2+\alpha +20\right) \text{bc}^2+\left(20 \alpha ^2-69\right) \text{bc}+9\right)\right)}-(16 \alpha +3) \text{bc}}{(2 \text{bc}-1) \left(\alpha +2 (\alpha -1) \text{bc}^2-3 \alpha  \text{bc}\right)},\\   \frac{-16 (\alpha -1) \text{bc}^4+(40 \alpha -26) \text{bc}^3+(13-32 \alpha ) \text{bc}^2+\sqrt{-\left((\text{bc}-1) \text{bc}^2 \left(-4 \alpha ^2+4 \left(4 \alpha ^2-4 \alpha -25\right) \text{bc}^3+8 \left(-4 \alpha ^2+\alpha +20\right) \text{bc}^2+\left(20 \alpha ^2-69\right) \text{bc}+9\right)\right)}+(8 \alpha -3) \text{bc}}{2 (\text{bc}-1) \text{bc} (2 \text{bc}-1) \left(\alpha +2 (\alpha -1) \text{bc}^2-3 \alpha  \text{bc}\right)},\\ \frac{16 (\alpha -1) \text{bc}^4+(26-40 \alpha ) \text{bc}^3+(32 \alpha -13) \text{bc}^2-\sqrt{-\left((\text{bc}-1) \text{bc}^2 \left(-4 \alpha ^2+4 \left(4 \alpha ^2-4 \alpha -25\right) \text{bc}^3+8 \left(-4 \alpha ^2+\alpha +20\right) \text{bc}^2+\left(20 \alpha ^2-69\right) \text{bc}+9\right)\right)}+(3-8 \alpha ) \text{bc}}{2 (\text{bc}-1) (2 \text{bc}-1) \left(\alpha +2 (\alpha -1) \text{bc}^2-3 \alpha  \text{bc}\right)}, 0 $\Big{)}, $P_9(\Omega_m)$= $
\frac{\text{bc} (2 \alpha +\text{bc} (-6 \alpha +(4 \alpha +6) \text{bc}-13)+3)-\sqrt{\text{bc}^2 \left(-4 \alpha ^2 \left(2 \text{bc}^2-3 \text{bc}+1\right)^2+8 \alpha  (\text{bc}-1) (2 \text{bc}-1) \text{bc}^2+\left(10 \text{bc}^2-13 \text{bc}+3\right)^2\right)}}{2 \text{bc} \left(\alpha +2 (\alpha -1) \text{bc}^2-3 \alpha  \text{bc}\right)}
$.\\
 $P_9(\omega_{\rm tot. })$=$\frac{
\alpha + \sqrt{\text{bc}^2 \left(-4 \alpha^2 \left(2 \text{bc}^2 - 3 \text{bc} + 1\right)^2 
+ 8 \alpha (\text{bc} - 1) (2 \text{bc} - 1) \text{bc}^2 
+ \left(10 \text{bc}^2 - 13 \text{bc} + 3\right)^2 \right)} 
+ \text{bc} \left(2 \alpha + \text{bc} \left(-19 \alpha - 4 \text{bc} \left(-7 \alpha + 3 (\alpha - 1) \text{bc} + 5\right) + 11 \right) - 3\right)
}{
3 (\text{bc} - 1) (2 \text{bc} - 1) \left(\alpha + 2 (\alpha - 1) \text{bc}^2 - 3 \alpha \text{bc}\right)
}$.}
\vspace{1cm}

In the preceding section, we have already established the autonomous system and are now prepared to identify the fixed points directly from the autonomous system equations eqs. \eqref{eq:d1}, \eqref{eq:d2}, \eqref{eq:d3}, \eqref{eq:d4}.   In this case, we have found a total of ten viable fixed points and they can be categorized into two scenarios: (i) Independent of curvature-matter coupling ($\alpha$) and model parameter ($bc$), (ii) Dependent of coupling and model parameter term. In the absence of coupling and model parameters, this interacting theory can come down to the non-interacting case with six critical points. We have listed down a total of ten viable critical points (finite and real). Parallelly, we have also found the critical matter density and total EOS parameter, which we also mentioned in the above tab. \ref{tab:1}. Below, we have provided a detailed analysis of the critical points.

\begin{figure}[H]
\centerline{$\begin{array}{cc}
\includegraphics[width=0.4\textwidth]{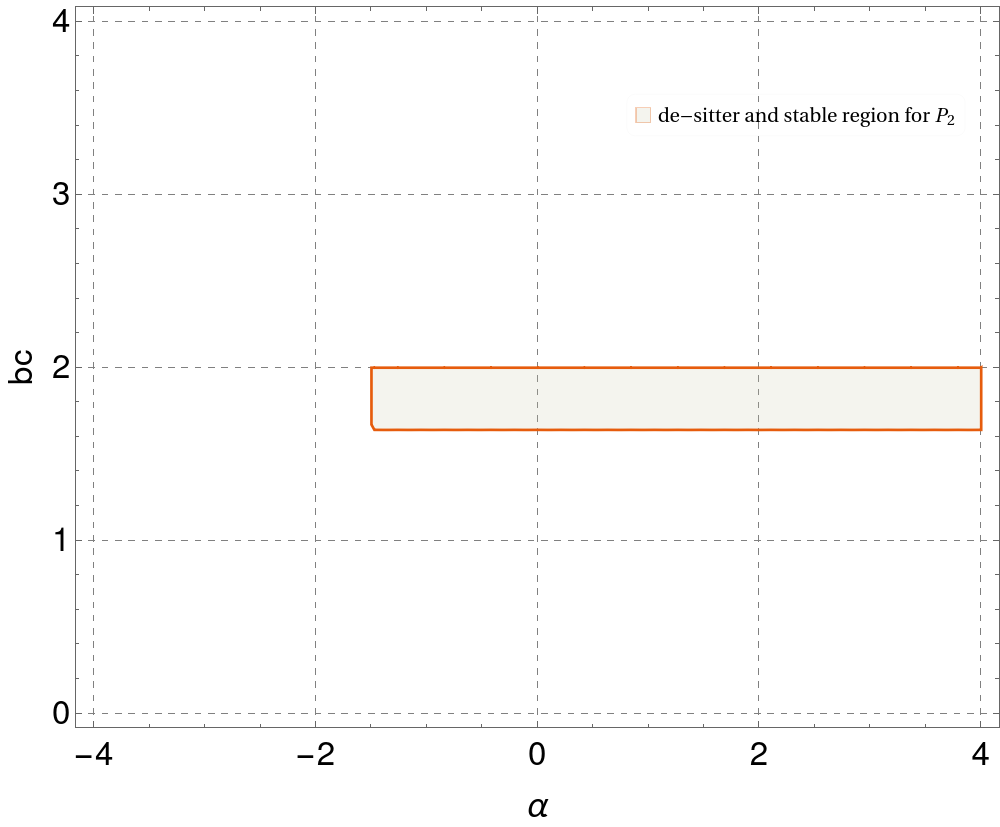}\quad \quad &
\includegraphics[width=0.48\textwidth]{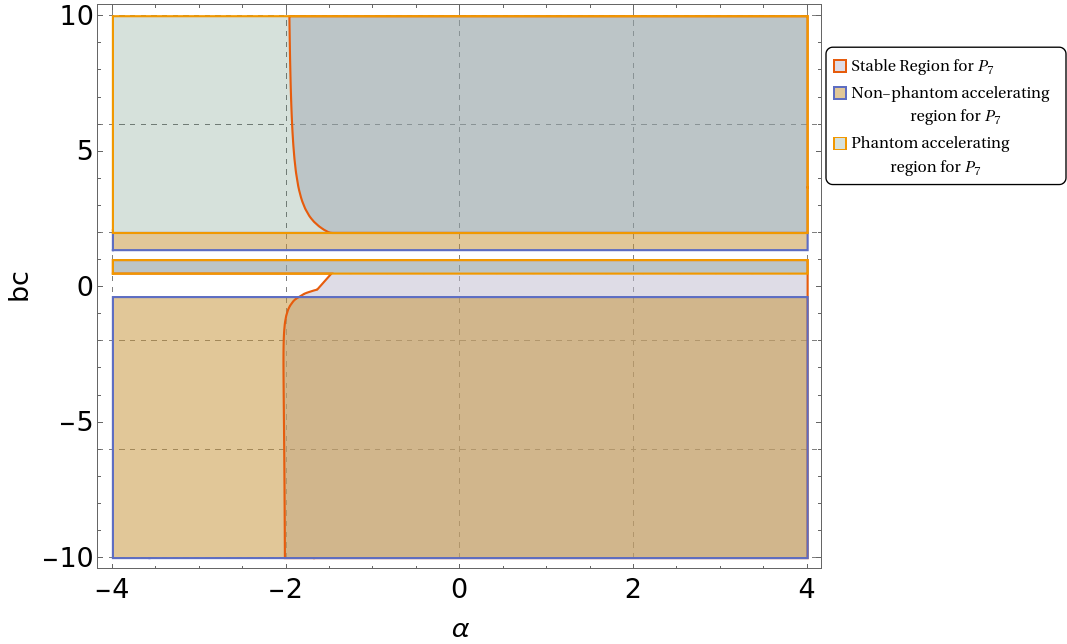}
\end{array}$}
\caption{Left Panel: Parameter space for generalised $\Lambda$CDM model at $P_2$, 
Right panel: Parameter space for generalised $\Lambda$CDM model at $P_7$.} 
\label{fig:1a}
\end{figure}

\begin{itemize}
\item $P_1$ Point: This fixed point is independent of any model parameter. In the context of this specific critical point, the dark matter density remains consistently at zero, and there is no evidence of late-time cosmic acceleration, regardless of the positive one-third total EOS parameter value.  This critical point represents a radiation-dominated epoch, characterized by a vanishing coincidence parameter. The absence of critical  matter density at this point implies that the universe is entirely governed by radiation, with no contribution from the matter or dark energy sectors. Such a point typically describes an early-time cosmological phase preceding matter or dark energy domination.

\item $P_2$ Point: The characteristics of this specific equilibrium point remain unaffected by model parameters. At this point, the dark matter density consistently remains at zero, and the EOS parameter is fixed at $-1$. This point represents a late-time accelerating de-Sitter solution dominated by dark energy. For $1.65 < bc < 2$ and $\alpha> -1.5$, this point shows stable behavior, which is depicted in the left panel of fig.\ref{fig:1a}.  This critical point corresponds to a de Sitter-like dark energy-dominated phase, where the cosmic expansion is entirely driven by dark energy. The coincidence parameter $( r_{\rm mc} = 0 )$ indicates that the critical matter density vanishes, confirming the absence of any matter contribution. Such a configuration typically represents a late-time attractor associated with accelerated expansion and asymptotic approach to a cosmological constant–dominated universe.

\item $P_{3\pm}$ Points: This specific set of critical points is expressed in terms of the coupling parameter $(\alpha)$. However, at these radiation-dominated critical points, the dark matter density remains unphysical, making them irrelevant to the scope of our study.  This critical point yields a constant cosmic coincidence parameter of the form
$r_{\rm mc} = \frac{-3 \pm 2\alpha + \sqrt{9 - 4\alpha^2}}{2\alpha}$,
suggesting a potential scaling behavior. However, the associated critical matter density at this point is unphysical—either negative or vanishing in regimes where it should be positive—rendering the solution cosmologically non-viable. As a result, despite its mathematical structure resembling a scaling solution, this point is excluded from physical consideration in the present analysis.

\item $P_{4\pm}$ Points: The critical points dependent on the coupling parameter exhibit an unphysical value for the critical dark matter density, despite representing a de-Sitter type solution.   The same expression for the coincidence parameter arises at these critical points; however, due to their unphysical nature, these points are also excluded from our analysis.

\item $P_{5\pm}$ Points: These critical points also exhibit $(\alpha)$-dependent behavior in a radiation-dominated scenario. However, the unphysical critical dark matter density makes these points irrelevant for consideration. In this case, the cosmic coincidence parameter is given by $r_{\rm mc} = \frac{1 - 2\alpha \pm \sqrt{1 - 4(-1 + \alpha)\alpha}}{2\alpha}$, which suggests the possibility of a scaling solution depending on the value of the coupling parameter \( \alpha \). However, the corresponding critical matter density associated with this point is unphysical—either negative or non-real within the relevant parameter space. Consequently, despite the form of $( r_{\rm mc} )$ indicating a mathematically admissible solution, this point is excluded from physical consideration in the cosmological context.

\item $P_{6}$ Point: This point represents a radiation-dominated scenario with pure kinetic dominance. No stable acceleration is observed at this particular critical point.   This critical point corresponds to a radiation-dominated epoch with a vanishing coincidence parameter, indicating zero critical  matter density. The universe is fully governed by radiation at this stage, typically representing an early-time cosmological phase preceding matter or dark energy domination.

\item $P_{7}$ Point: This critical point is dependent on the model parameters, with the critical matter density at this point being zero. The stable and accelerating regions are shown in the right panel of fig.~\ref{fig:1a}. For $( bc < -0.5 )$ and $( \alpha > -2 )$, we have identified a common region where a benchmark point can be selected that provides a stable and  non-phantom type accelerating late-time solution. We also indicated the phantom-type accelerating region in the same plot; however, this region corresponds to an unstable accelerating solution. This critical point corresponds to a dark energy-dominated phase, characterized by a vanishing coincidence parameter  and zero critical matter density. In this regime, the dynamics of the universe are entirely driven by dark energy, with no contribution from the matter sector.

\item $P_{8}$ \& $P_{9}$ Points:  Both critical points have been expressed in terms of the model parameter $(bc)$ and the coupling parameter $(\alpha)$. The critical matter density and the total equation of state (EoS) parameter corresponding to these points have also been determined in terms of the model parameter. However, these points are unable to simultaneously exhibit both a stable and an accelerating solution.   Value of coincidence parameter is 
{\tiny  \begin{align*}
 r_{\rm mc} = &\frac{\pm
\sqrt{
bc^2 \left[
(3 - 13\,bc + 10\,bc^2)^2
+ 8(-1 + bc)\,bc^2\,(-1 + 2\,bc)\,\alpha
- 4(1 - 3\,bc + 2\,bc^2)^2\,\alpha^2
\right]
} 
- (1 - bc)\,bc\,\left(-3 - 2\alpha + 2\,bc\,(5 + 2\alpha)\right)
}{
2(1 - bc)\,bc\,(-1 + 2\,bc)\,\alpha
} 
\end{align*}}
This expression suggests a potential scaling solution, depending on the values of the parameters \( \alpha \) and \( bc \). The presence of the \( \pm \) sign indicates two possible branches, of which only one may lead to a physically viable, positive value of \( r_{\rm mc} \). The reality and positivity of this expression are essential for it to represent a meaningful cosmological scenario. Otherwise, the critical point may be discarded due to unphysical behavior such as negative or non-real matter density.

 \item $P_{10}$ Point: This point displays a saddle-like behavior, lacking both stability and cosmic acceleration for any parameter values.  In the limiting case as \( bc \to 1 \), the dynamics become purely radiation-dominated, with a vanishing coincidence parameter.
\end{itemize}

As our study primarily focuses on the late-time phase of cosmic evolution, we previously discussed an interaction between the curvature and dark-matter sectors while neglecting the radiation sector. Consequently, in the phase space diagram, we set $X_4 = 0$ and analyze the phase space behavior of the 3-D system spanned by the variables $X_1$, $X_2$, and $X_3$. In the present case, the phase space is unbounded, as $-\infty < X_1 < \infty$, $-\infty < X_2 < \infty$, and $-\infty < X_3 < \infty$. To represent the phase space within a bounded region, we apply a transformation to the phase space variables. These transformations connect the conventional phase space variables $(X_1, X_2, X_3)$ to a new set of variables $(X, Y, Z)$, defined as:  
\[
X = \tan^{-1}(X_1), \quad Y = \tan^{-1}(X_2), \quad Z = \tan^{-1}(X_3).
\]  
This mapping transforms the unbounded phase space into a bounded region with:  
\[
-\frac{\pi}{2} < X < \frac{\pi}{2}, \quad -\frac{\pi}{2} < Y < \frac{\pi}{2}, \quad -\frac{\pi}{2} < Z < \frac{\pi}{2}.
\]

The dynamical variables $X_1, X_2, X_3$ are unbounded in general. However, from matter energy density criterion in eqn.\eqref{eq:d7}, it follows that  
$$  
0 \leq \Omega_m \leq 1,  
$$  
which translates to  
\begin{eqnarray}
 0 \leq (X_1 + X_2 + X_3) \leq 1
\label{eq:e1}   
\end{eqnarray}

\vspace{2cm}

Here, the radiation component $X_4$ is assumed to be zero. To reformulate this in the context of a transformed phase-space representation, we use a new set of variables. In this framework, the condition can be expressed as,  
\begin{eqnarray}
0 \leq (X+Y+Z) \leq \frac{\pi}{4}
\label{eq:e2}
\end{eqnarray}
By implementing these conditions, we plot the phase space for both types of interacting models. Within these constraints, stable critical points and other viable fixed points are identified. Crucially, imposing this specific constraint ensures that the analysis remains confined within the bounded region of the phase space. Consequently, this eliminates the need to investigate critical points at infinity, which typically arise due to the unbounded nature of the original variables. Such points, while mathematically possible, often correspond to unrealistic or unphysical solutions. By restricting the analysis to the bounded region, the physical viability of the models is better ensured, and the study becomes more computationally tractable.\\

\begin{figure}[H]
\centerline{$\begin{array}{cc}
\includegraphics[width=0.4\textwidth]{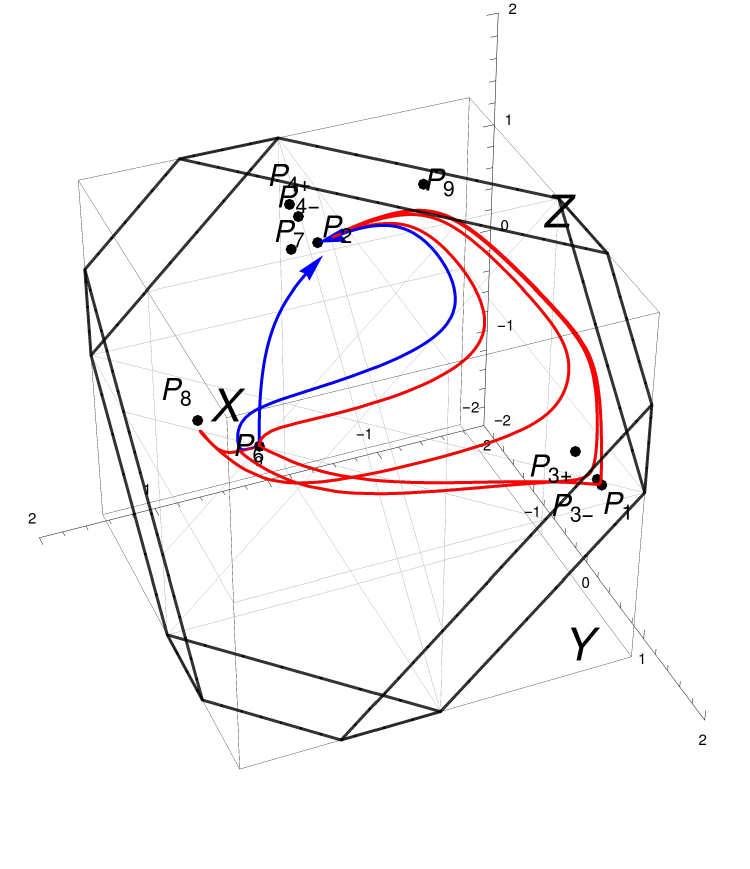}\quad \quad &
\includegraphics[width=0.4\textwidth]{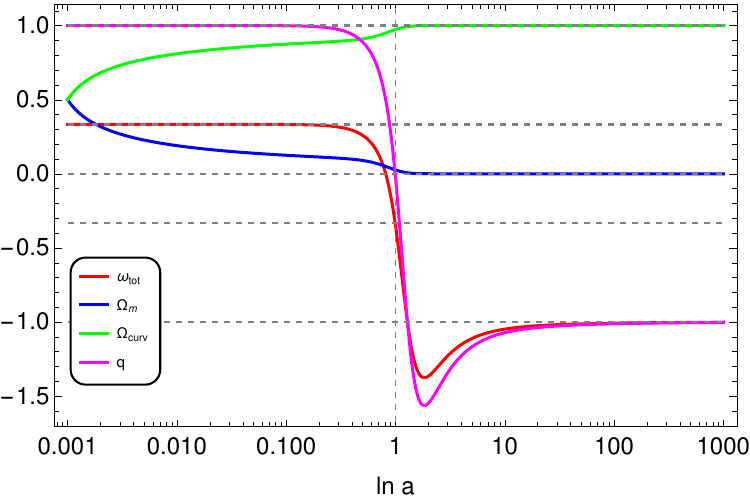}
\end{array}$}
\caption{Left Panel: Phase-space trajectories for generalised $\Lambda$CDM model for $bc=1.75, \alpha=1$ (bold black line in the phase-space corresponding to the constraint of $0 \leq X+Y+Z \leq \frac{\pi}{4}$), Right panel: Evolution plots of cosmological parameters for $bc=1.75, \alpha=1$. Both plots are corresponding to $P_2$ (model parameters independent) as a stable point and choose $X_4=0$ (absence of radiation component).} 
\label{fig:1b}
\end{figure}

To analyze the phase-space behavior of this system, we have divided the study into two cases. In the first case, we consider the critical point $P_2$ as a model-independent attractor, while in the second case, $P_7$ is treated as a model-dependent attractor. To proceed with the analysis, we select benchmark values for $(bc, \alpha)$ such that they fall within the parameter space for both cases, ensuring a   stable non-phantom type of late-time cosmic accelerating solution. To illustrate the attractor behavior of the critical point $P_2$ in the left panel of fig.~\ref{fig:1b}, we have selected the benchmark values $bc = 1.75$ and $\alpha = 1$. In the phase-space trajectories, only the points yielding real values are shown, while the other critical points are omitted. The red-line trajectories correspond to repeller-type behavior, while the blue lines represent attractor-type behavior. From the phase-space plot, we observe that all red-line trajectories are repelled from the critical points $P_1$, $P_{3\pm}$, $P_6$, $P_8$, and $P_9$, indicating that these points exhibit saddle or unstable behavior. In contrast, the critical point $P_2$ demonstrates stable behavior, attracting all repelling trajectories towards it. Additionally, some trajectories (marked in blue) are shown directly converging to the point $P_2$, thereby establishing this de Sitter point as an attractor.\\

In the right panel of fig.~\ref{fig:1b}, we present the evolutionary dynamics of cosmological parameters as a function of the logarithm of the FLRW scale factor $a$ for the model parameters $bc = 1.75, \alpha =1$. This plot incorporates parameters such as the total equation of state (EoS) parameter, critical dark matter density, curvature density, and the deceleration parameter, showcasing their behavior within the framework of this interacting model. Magnitude of the total EoS parameter begins at $\frac{1}{3}$, indicating an initial radiation-dominated phase where curvature and matter densities are equal. Over time, the dark matter density decreases and asymptotically approaches zero, while the curvature density increases and tends toward unity. This behavior suggests that the curvature component effectively mimics the role of dark energy in the late-time evolution of the universe. Furthermore, the evolution plot reveals energy transfer from the dark matter sector to the curvature sector.  Variations of the total EoS and deceleration parameters are depicted across different epochs, transitioning from radiation domination to matter domination, and ultimately culminating in late-time dark energy domination extending into the distant future. Notably, the total EoS parameter exhibits a dynamical behavior, crossing into the phantom region before stabilizing at the de Sitter point. Additionally, the deceleration parameter undergoes a transition from a positive unity value during the radiation epoch to negative unity in the dark energy-dominated era.\\

\begin{figure}[H]
\centerline{$\begin{array}{cc}
\includegraphics[width=0.4\textwidth]{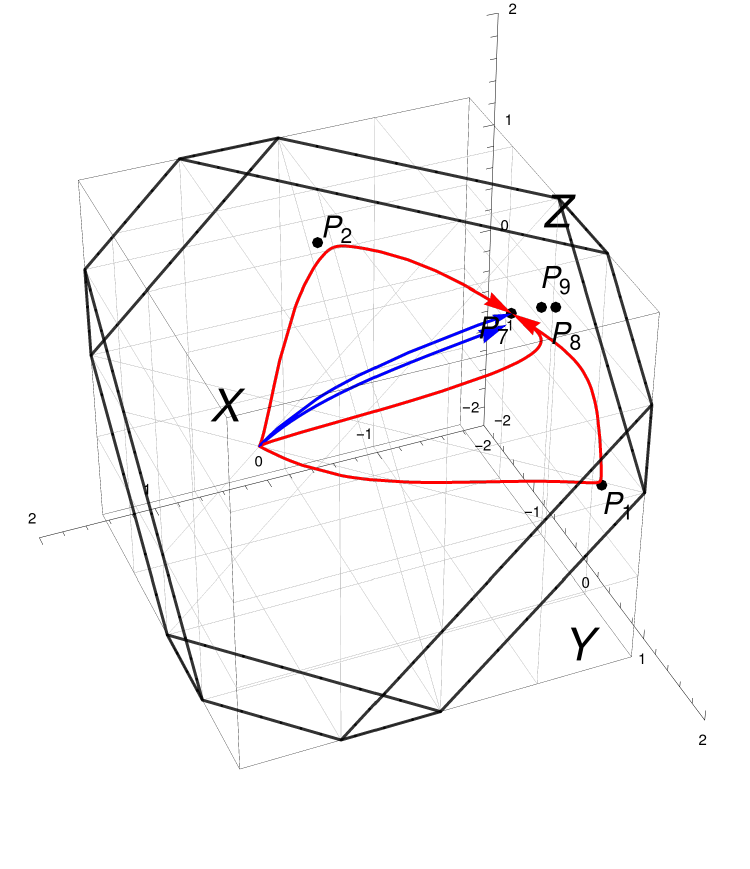}\quad \quad &
\includegraphics[width=0.4\textwidth]{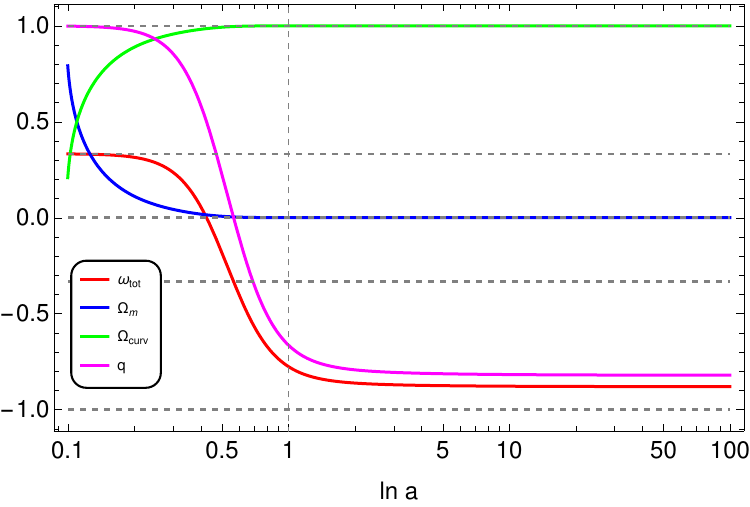}
\end{array}$}
\caption{Left Panel: Phase-space trajectories for generalised $\Lambda$CDM model for $bc=-3, \alpha=2$ (bold black line in the phase-space corresponding to the constraint of $0 \leq X+Y+Z \leq \frac{\pi}{4}$), 
Right panel: Evolution plots of cosmological parameters for 
 $bc=-3, \alpha=2$. Both plots are corresponding to $P_7$ (model parameters dependent) as a stable point and choose $X_4=0$ (absence of radiation component).} 
\label{fig:1c}
\end{figure}

Similarly, we have chosen another benchmark to demonstrate the stable behavior of the model-parameter-dependent critical points, which arise specifically in the case of an interacting curvature-matter scenario. In the left panel of fig.~\ref{fig:1c}, we have selected a benchmark value of $(bc = -3, \alpha = 2)$, and for this selection, we refer to fig.~\ref{fig:1a}, where the critical point $P_7$ exhibits both stability and   non-phantom type of accelerating behavior. Using this benchmark, we have included only those critical points in the phase-space plot that yield real values. As in the previous case, two types of trajectories are present: blue lines (representing attractors) and red lines (representing repellers). All trajectories converge toward $P_7$, confirming it as an attractor. Depending on the nature of the coupling and the model parameters, the behavior of critical points transitions between repeller and attractor. This change in the nature of the critical points is particularly observed for $P_2$ and $P_7$. In the right panel of fig.~\ref{fig:1c}, we depict the evolution of cosmological parameters for the model with $bc = -3$ and $\alpha = 2$, similar to the previous case. The plot includes the total equation of state (EoS) parameter, critical dark matter and curvature densities, and the deceleration parameter. The total EoS parameter begins at $\frac{1}{3}$, indicating a radiation-dominated phase with a dominant contribution from the dark matter sector over the curvature component. Over time, the dark matter density decreases toward zero, while the curvature density rises to unity, mimicking dark energy at late times. An energy transfer from the dark matter sector to the curvature sector is evident. The evolution of the total EoS and deceleration parameters reveal transitions from radiation domination to matter domination and ultimately to a dark energy-dominated era. The total EoS parameter crosses the matter-dominated phase, enters the dark energy regime with $\omega_{\rm tot} = -\frac{1}{3}$, and stabilizes near $-0.9$. This interacting model successfully demonstrates a curvature-driven, non-phantom dark energy scenario at late times. This interacting model successfully captures a comprehensive evolutionary trajectory, encompassing all major phases of the universe's history.

\subsection{Fixed point stability analysis  in power-law  model}

By utlising the autonomous equations mentioned in the last section with eqs. \eqref{eq:d1}, \eqref{eq:d2}, \eqref{eq:d3}, \eqref{eq:d4}, we have found a total of eight critical points for the interacting power-law type $f(R)$ model. Like the previous case, here we also categorised the fixed points based on their nature. Presence of interaction gives rise the new fixed points, which we will discuss onwards. By listing down only real and finite critical points in the table along with critical dark matter density and total EoS parameter in tab. \ref{tab:2}.

\begin{table}[H]
\centering
\begin{tabular}{|c|c|c|c|}
\hline
 Critical  & \multirow{2}{*}{($X_1,X_2,X_3,X_4$)} & \multirow{2}{*}{$\Omega_{\rm m} $} & \multirow{2}{*}{$\omega_{\rm tot}$} \\
points & & &  \\
\hline\hline
$P_{1}$ & ($-4$ , 5 , 0 , 0) & 0 & $\frac13$  \\
\hline
$P_{2}$ & (0 , $-1$ , 2 , 0) & 0 & $-1$  \\
\hline
\multirow{2}{*}{$P_{3 \pm}$} & \multirow{2}{*}{$\left( -4 , \frac{-3  \mp  8\alpha + \sqrt{9-4\alpha^2} }{2\alpha} , 0 , 0 \right)$} & \multirow{2}{*}{$\frac{\mp \sqrt{9-4 \alpha ^2}+2 \alpha \mp 3}{2 \alpha }$} & \multirow{2}{*}{$\frac13$}  \\
&&&\\
\hline
\multirow{2}{*}{$P_{4 \pm}$} & \multirow{2}{*}{$\left(0 , \frac{- 3 \mp  4\alpha + \sqrt{9-4\alpha^2}}{2\alpha} , 2 , 0\right)$} & \multirow{2}{*}{$\mp \frac{\sqrt{9-4 \alpha ^2}+2 \alpha \mp 3}{2 \alpha }$} & \multirow{2}{*}{$-1$} \\
&&&\\
\hline
\multirow{2}{*}{$P_{5}$} & \multirow{2}{*}{$\left(-\frac{2 (n-2)}{2 n-1}, \frac{5-4 n}{2 n^2-3 n+1}, \frac{n (4 n-5)}{2 n^2-3 n+1},0 \right)$} & \multirow{2}{*}{$ 0$} & \multirow{2}{*}{$\frac{-6 n^2+7 n+1}{6 n^2-9 n+3}$} \\
&&&\\
\hline
\multirow{2}{*}{$P_6$} & \multirow{2}{*}{$P_6(X_1, X_2, X_3, •••••••••••0$)} & \multirow{2}{*}{$ P_6 (\Omega_m) $} & \multirow{2}{*}{$P_6(\omega_{\rm tot.})$} \\
&&&\\
\hline
\multirow{2}{*}{$P_7$} & \multirow{2}{*}{$P_7(X_1, X_2, X_3, 0$)} & \multirow{2}{*}{$ P_7 (\Omega_m) $} & \multirow{2}{*}{$P_7(\omega_{\rm tot.})$} \\
&&&\\
\hline

\multirow{2}{*}{$P_{8}$} & \multirow{2}{*}{\scriptsize $\left( \frac{4 (-1 + n)}{n}, \frac{-(2 (-1 + n)}{n^2},\frac{2 (-1 + n)}{n}, \frac{(-2 + 8 n - 5 n^2)}{n^2} \right)$} & \multirow{2}{*}{$ 0 $} & \multirow{2}{*}{$-1+\frac{4}{3n}$} \\
&&&\\
\hline

\end{tabular}
\caption{List of critical points of the power-law model with corresponding $\Omega_{\rm m}$ and $\omega_{\rm tot}$. Details of the critical points $P_6$ and $P_7$ are provided below.}

\label{tab:2}
\end{table}
\vspace{1cm}

 Where,
{\scriptsize
$P_6(X_1, X_2, X_3, 0)$ =\\ $\Bigg( 
\frac{
4\alpha - 2(4\alpha + 1)n^3 + 5(4\alpha + 1)n^2
}{
(2n-1) 
\left(
\alpha + 2(\alpha - 1)n^2 - 3\alpha n
\right)
}  
+ \frac{
\sqrt{
-\bigg(
(n-1)n^2 \Big(
-4\alpha^2 + 4(4\alpha^2 - 4\alpha - 25)n^3 
+ 8(-4\alpha^2 + \alpha + 20)n^2 
+ (20\alpha^2 - 69)n + 9
\Big)
\bigg)} - (16\alpha + 3)n
}{
(2n-1) 
\left(
\alpha + 2(\alpha - 1)n^2 - 3\alpha n
\right)
},\\
\frac{
-16(\alpha - 1)n^4 + (40\alpha - 26)n^3 + (13 - 32\alpha)n^2
}{
2n 
\left(
2n^2 - 3n + 1
\right) 
\left(
\alpha + 2(\alpha - 1)n^2 - 3\alpha n
\right)
}  
+ \frac{
\sqrt{
-\bigg(
(n-1)n^2 \Big(
-4\alpha^2 + 4(4\alpha^2 - 4\alpha - 25)n^3 
+ 8(-4\alpha^2 + \alpha + 20)n^2 
+ (20\alpha^2 - 69)n + 9
\Big)
\bigg)} + (8\alpha - 3)n
}{
2n 
\left(
2n^2 - 3n + 1
\right) 
\left(
\alpha + 2(\alpha - 1)n^2 - 3\alpha n
\right)
},\\
\frac{
16(\alpha - 1)n^4 + (26 - 40\alpha)n^3 + (32\alpha - 13)n^2
}{
2 
\left(
2n^2 - 3n + 1
\right) 
\left(
\alpha + 2(\alpha - 1)n^2 - 3\alpha n
\right)
} 
- \frac{
\sqrt{
-\bigg(
(n-1)n^2 \Big(
-4\alpha^2 + 4(4\alpha^2 - 4\alpha - 25)n^3 
+ 8(-4\alpha^2 + \alpha + 20)n^2 
+ (20\alpha^2 - 69)n + 9
\Big)
\bigg)} + (3 - 8\alpha)n
}{
2 
\left(
2n^2 - 3n + 1
\right) 
\left(
\alpha + 2(\alpha - 1)n^2 - 3\alpha n
\right)
},  0
\Bigg),
$
}

{\scriptsize
$P_6(\Omega_m) $ = $ \frac{
(4\alpha + 6)n^3 - (6\alpha + 13)n^2
}{
2n \left(\alpha + 2(\alpha - 1)n^2 - 3\alpha n\right)
}   
- \frac{
\sqrt{
-\left((n-1)n^2 
\left(
-4\alpha^2 + 4(4\alpha^2 - 4\alpha - 25)n^3 + 8(-4\alpha^2 + \alpha + 20)n^2 
+ (20\alpha^2 - 69)n + 9
\right)
\right)} + (2\alpha + 3)n
}{
2n \left(\alpha + 2(\alpha - 1)n^2 - 3\alpha n\right)
},$\\
$P_6(\omega_{\text{tot.}})$  =\\ $\frac{
\alpha - 12(\alpha - 1)n^4 + 4(7\alpha - 5)n^3 + (11 - 19\alpha)n^2
}{
3 \left(2n^2 - 3n + 1\right) \left(\alpha + 2(\alpha - 1)n^2 - 3\alpha n\right)
}  
+ \frac{
\sqrt{
-\left((n-1)n^2 
\left(•••••••••••
-4\alpha^2 + 4(4\alpha^2 - 4\alpha - 25)n^3 + 8(-4\alpha^2 + \alpha + 20)n^2 
+ (20\alpha^2 - 69)n + 9
\right)
\right)} + (2\alpha - 3)n
}{
3 \left(2n^2 - 3n + 1\right) \left(\alpha + 2(\alpha - 1)n^2 - 3\alpha n\right)}$}\\.

\vspace{1cm}

{\scriptsize 
 
$P_7(X_1, X_2, X_3, 0)$  =\\ $\Bigg( 
-\frac{.
-4\alpha + (8\alpha + 2)n^3 - 5(4\alpha + 1)n^2
}{
(2n-1)
\left(
\alpha + 2(\alpha - 1)n^2 - 3\alpha n
\right)
}  
+ \frac{
\sqrt{
-\left(
(n-1)n^2 \left(
-4\alpha^2 + 4(4\alpha^2 - 4\alpha - 25)n^3 
+ 8(-4\alpha^2 + \alpha + 20)n^2 
+ (20\alpha^2 - 69)n + 9
\right)
\right)} + (16\alpha + 3)n
}{
(2n-1) 
\left(
\alpha + 2(\alpha - 1)n^2 - 3\alpha n
\right)
}, \\
\frac{
n^2(13 - 32\alpha) - 16n^4(\alpha - 1) + n(3 - 8\alpha)
}{
2n(1 - 3n + 2n^2)
\left(
2n^2(\alpha - 1) + \alpha - 3\alpha n
\right)
} 
+ \frac{
n^3(-26 + 40\alpha) - \sqrt{
-\left(
(n-1)n^2 \left(
9 - 4\alpha^2 + 8n^2(20 + \alpha - 4\alpha^2) 
+ 4n^3(-25 - 4\alpha + 4\alpha^2) + n(-69 + 20\alpha^2)
\right)
\right)
}
}{
2n(1 - 3n + 2n^2)
\left(
2n^2(\alpha - 1) + \alpha - 3\alpha n
\right)
}, \\
\frac{
n^3(26 - 40\alpha) + n(3 - 8\alpha) + 16n^4(\alpha - 1)
}{
2(1 - 3n + 2n^2)
\left(
2n^2(\alpha - 1) + \alpha - 3\alpha n
\right)
}  
+ \frac{
n^2(-13 + 32\alpha) + \sqrt{
-\left(
(n-1)n^2 \left(
9 - 4\alpha^2 + 8n^2(20 + \alpha - 4\alpha^2) 
+ 4n^3(-25 - 4\alpha + 4\alpha^2) + n(-69 + 20\alpha^2)
\right)
\right)
}
}{
2(1 - 3n + 2n^2)
\left(
2n^2(\alpha - 1) + \alpha - 3\alpha n
\right)
},   0
\Bigg).$
}
  
{\scriptsize

$P_7(\Omega_m)$  = $\frac{
(4\alpha + 6)n^3 - (6\alpha + 13)n^2
}{
2n \left(\alpha + 2(\alpha - 1)n^2 - 3\alpha n\right)
}  
+ \frac{
\sqrt{
-\left((n-1)n^2 
\left(
-4\alpha^2 + 4(4\alpha^2 - 4\alpha - 25)n^3 + 8(-4\alpha^2 + \alpha + 20)n^2 
+ (20\alpha^2 - 69)n + 9
\right)
\right)} + (2\alpha + 3)n
}{
2n \left(\alpha + 2(\alpha - 1)n^2 - 3\alpha n\right)
},$\\
$P_7(\omega_{\text{tot.}})$ =\\ $\frac{
\alpha - 12(\alpha - 1)n^4 + 4(7\alpha - 5)n^3 + (11 - 19\alpha)n^2
}{
3 \left(2n^2 - 3n + 1\right) \left(\alpha + 2(\alpha - 1)n^2 - 3\alpha n\right)
}  
- \frac{
\sqrt{
-\left((n-1)n^2 
\left(
-4\alpha^2 + 4(4\alpha^2 - 4\alpha - 25)n^3 + 8(-4\alpha^2 + \alpha + 20)n^2 
+ (20\alpha^2 - 69)n + 9
\right)
\right)} + (2\alpha - 3)n
}{
3 \left(2n^2 - 3n + 1\right) \left(\alpha + 2(\alpha - 1)n^2 - 3\alpha n\right)
}$
}\\

A detailed analysis of the critical points is provided below.

\begin{figure}[H]
\centerline{$\begin{array}{cc}
\includegraphics[width=0.4\textwidth]{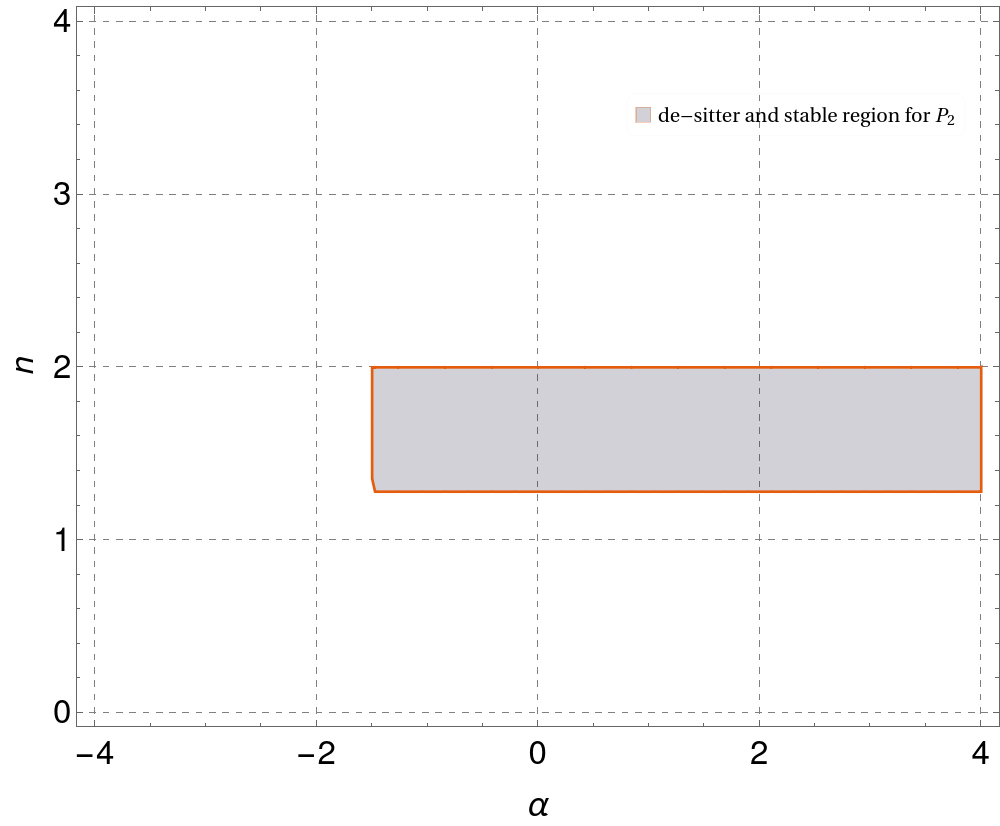}\quad \quad &
\includegraphics[width=0.48\textwidth]{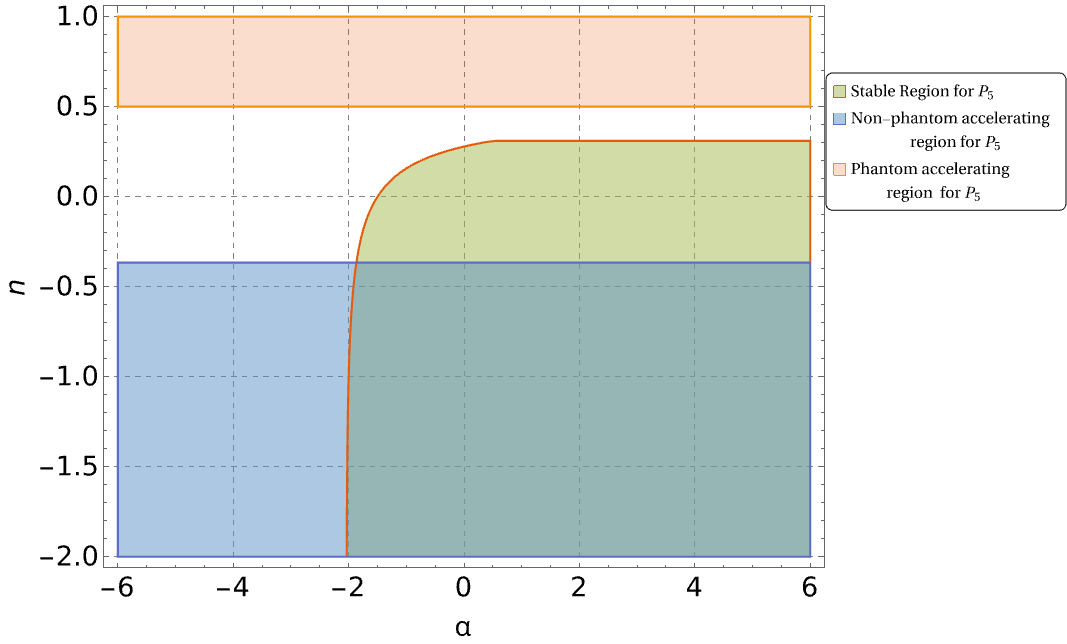}
\end{array}$}
\caption{Left Panel: Parameter space for power-law model at $P_2$, 
Right panel: Parameter space for power-law model at $P_5$.} 
\label{fig:2a}
\end{figure}

 \begin{itemize}
\item $P_1$ Point:  This model-independent critical point, characterized by a positive total EoS parameter of $\frac{1}{3}$, arises in all cases, regardless of the presence or absence of interaction. The critical matter density at this point is always zero.    Since the critical matter density vanishes at this critical point, the cosmic coincidence parameter takes a value of zero.

\item $P_2$ Point: Nature of this fixed point is independent of the model parameters. At this point, the dark matter density is consistently zero, and the equation of state (EoS) parameter is fixed at $-1$. This point represents a late-time accelerating de-Sitter solution dominated by dark energy. For $1.3 < bc < 2$ and $\alpha > -1.4$, the point exhibits stable behavior, as shown in the left panel of fig.~\ref{fig:2a}.  This point is entirely dominated by dark energy, with a vanishing critical   matter density, yielding $r_{\rm mc} = 0$.

 \item $P_{3\pm}$ Points: This specific set of critical points is expressed in terms of the coupling parameter $(\alpha)$. However, at these radiation-dominated critical points, the dark matter density remains unphysical, making them irrelevant to the objectives of our study.  For this case, a scaling behavior arises when the cosmic coincidence parameter approaches a constant value, given as 
$r_{\rm mc} = \frac{-3 \pm 2\alpha + \sqrt{9 - 4\alpha^2}}{2\alpha}$, indicating that the critical energy densities of  matter and dark energy evolve in a fixed proportion. This corresponds to a genuine scaling solution, where neither component overwhelmingly dominates the cosmic dynamics. However, due to the emergence of an unphysical  critical  matter density at these points, they are excluded from our physical analysis.

\item $P_{4\pm}$ Points: The critical points dependent on the coupling parameter exhibit an unphysical value for the critical dark matter density, despite representing a de-Sitter type solution.   Although these points exhibit a scaling behavior similar to the previously discussed points \( P_{3\pm} \), they yield an unphysical critical matter density. Therefore, we exclude these cases from our analysis.

 \item $P_{5}$ Point: The dynamics of the power-law model under the influence of interactions are significantly influenced by this critical point. This fixed point depends on the model parameter $(n)$ and is characterized by the absence of a radiation component $(X_4 = 0)$. At this point, the critical   matter density is always zero. The total equation of state (EoS) parameter and the stability condition of this fixed point also depend on the model parameters. In fig.~\ref{fig:2a}, we illustrate the common region in the parameter space of $(n)$ and $(\alpha)$ where this critical point $(P_5)$ exhibits a stable,   non-phantom type accelerating condition. Phantom-type accelerating region is also highlighted in the same plot, though it corresponds to an unstable form of acceleration. From the overlapping region of stability and non-phantom acceleration, we select a benchmark point corresponding to specific values of $(n)$ and $(\alpha)$ to analyze the complete dynamics of this fixed point through phase-space trajectories and evolution plots.  The condition \( r_{\rm mc} = 0 \) corresponds to a dark energy-dominated solution, where the critical matter density vanishes entirely. In this regime, the cosmic dynamics are fully driven by dark energy, with no contribution from the matter sector at the critical point.

\item $P_{6}$ \& $P_{7}$ Points: These critical points exhibit a dependency on the model parameters in all aspects, including their coordinate values, critical dark matter density, and the total equation of state (EoS) parameter. However, this particular set of critical points is excluded because they simultaneously satisfy conditions for both stability and acceleration, which is inconsistent with the desired physical behavior.
   Value of coincidence parameter is 
{\scriptsize \begin{align*}
r_{\rm mc} & = \frac{1}{
2(1 - n)\,n\,(-1 + 2n)\,\alpha
}
\Bigg(\pm 
\sqrt{
n^2 \left[
(3 - 13\,n + 10\,n^2)^2
+ 8(-1 + n)\,n^2\,(-1 + 2n)\,\alpha
- 4(1 - 3\,n + 2\,n^2)^2\,\alpha^2
\right]
} \\
&\quad + (1 - n)\,n\,\left(-3 - 2\alpha + 2n(5 + 2\alpha)\right)
\Bigg)
\end{align*}}
This expression corresponds to a general scaling behavior in the system, with the value of \( r_{\rm mc} \) depending on both the coupling parameter \( \alpha \) and the parameter \( n \).

 \item $P_{8}$ Point: This point displays a saddle-like behavior, lacking both stability and cosmic acceleration for any parameter values.   The condition \( r_{\rm mc} = 0 \) directly corresponds to a scenario where the critical matter density vanishes. Within this viable range, the total equation of state parameter \( \omega_{\rm tot} \) remains greater than \( \frac{1}{3} \). However, in the limiting case as \( n \to 1 \), the dynamics become purely radiation-dominated. Outside this limit, no cosmologically interesting behavior is found in this region.

\end{itemize}

Similar to the previous model, our focus remains on the matter-dominated and late-time phases, as well as the interaction between matter and curvature, while neglecting the radiation-dominated phase. For a compact phase-space analysis, we restrict our attention to the three coordinates $(X_1, X_2, X_3)$ and their transformation into $(X, Y, Z)$ with the constraints mentioned in eqs. \eqref{eq:e1},\eqref{eq:e2}. We have selected a benchmark point for the model and coupling parameter values from the common allowable region that satisfies the stable-accelerating condition at the $P_5$ point. Only real-valued critical points are included in the phase-space analysis. The phase-space trajectories are drawn using this benchmark point because $P_5$ uniquely captures the effects of interaction.\\

\begin{figure}[H]
\centerline{$\begin{array}{cc}
\includegraphics[width=0.4\textwidth]{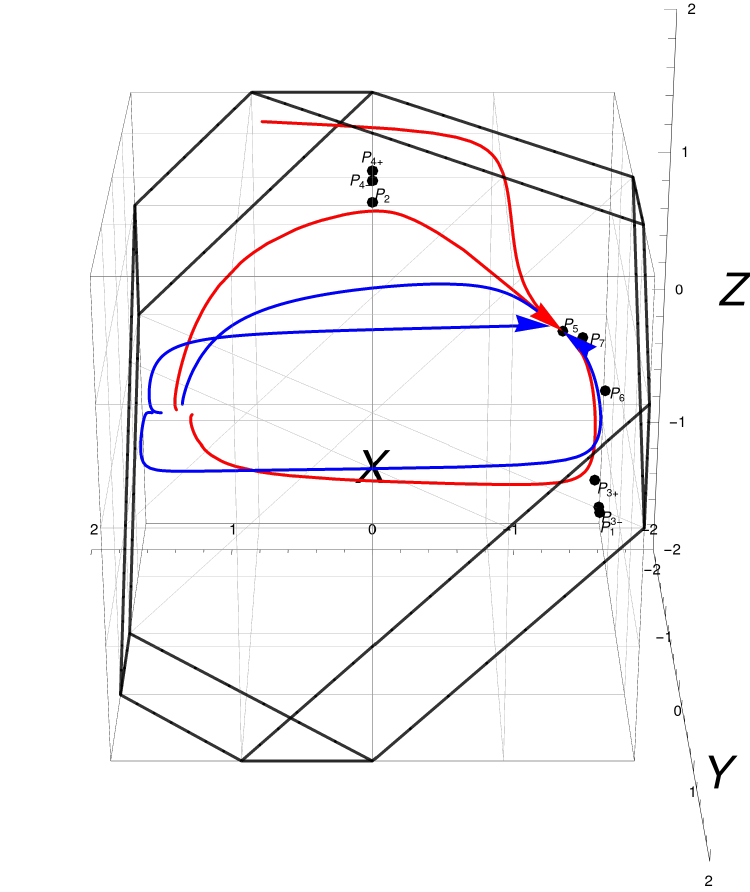}\quad \quad &
\includegraphics[width=0.4\textwidth]{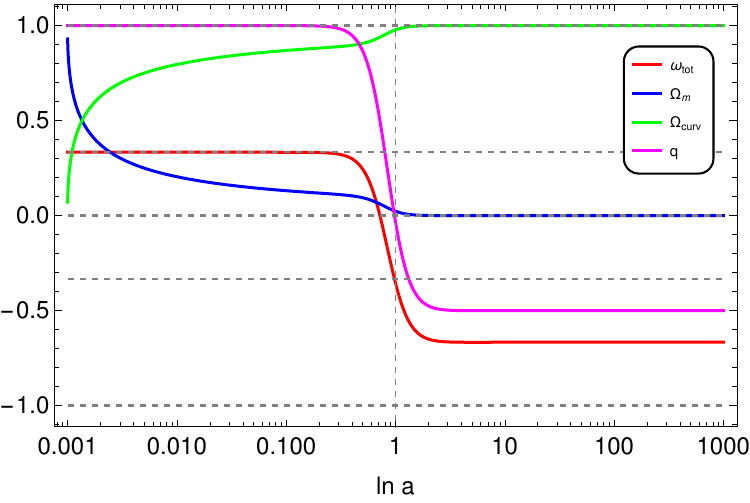}
\end{array}$}
\caption{Left Panel: Phase-space trajectories for generalised $\Lambda$CDM model for $n=-1, \alpha=1$ (bold black line in the phase-space corresponding to the constraint of $0 \leq X+Y+Z \leq \frac{\pi}{4}$), 
Right panel: Evolution plots of cosmological parameters for 
 $n=-1, \alpha=1$. Both plots correspond to $P_5$ (model parameters dependent) as a stable point and choose $X_4=0$ (absence of radiation component).} 
\label{fig:2b}
\end{figure}

In the left panel of fig.~\ref{fig:2b}, we present the phase-space trajectories to examine the dynamical evolution of the system. Similar to the previous scenario, two distinct types of trajectories are highlighted using red and blue to represent different dynamical behaviors. For this analysis, we proceed with specific model parameter values: $n = -1$ and $\alpha = 1$, chosen as the benchmark from the allowable region. Under these parameter values, only one fixed point exhibits the characteristics of an attractor, effectively drawing trajectories toward itself. In contrast, all other critical points repel the trajectories in their vicinity, confirming their nature as unstable points. Notably, the fixed point $P_2$, which shows different behavior under other parameter choices, turns into a repeller at this particular benchmark. The red trajectories are specifically marked to illustrate the saddle-like behavior of critical points where the trajectories neither converge toward nor diverge away uniformly but exhibit intermediate stability patterns. This analysis highlights the influence of parameter selection on the dynamical properties of the critical points and provides a clear visualization of the distinct roles played by attractors and repellers in the context of interacting curvature-matter scenario. \\

In the right panel of fig.~\ref{fig:2b}, we present the evolution of key cosmological parameters for the interacting power-law model with parameter values $n = -1$ and $\alpha = 1$. The plot tracks the dynamical evolution of the total equation of state (EoS) parameter $\omega_{\rm tot}$, the critical densities of dark matter and curvature, and the deceleration parameter $q$. At the outset, the magnitude of the total EoS parameter begins at $ \frac{1}{3}$, indicating a radiation-dominated phase. In this early era, the matter sector dominates over the curvature component. As time progresses, the dark matter density $ \Omega_{\rm m}$ gradually decreases, approaching zero, while the curvature density $\Omega_{\rm curv}$ rises toward unity, effectively mimicking a dark energy-like behavior at late times. This transition demonstrates a clear energy transfer from the dark matter sector to the curvature sector, a hallmark of the interacting model. The evolution of ($\omega_{\rm tot}$ and $q$) reveals a series of cosmological transitions. Initially, the universe transitions from a radiation-dominated phase to a matter-dominated phase. Later, it enters a dark energy-dominated era. During this progression, the total EoS parameter $\omega_{\rm tot}$ crosses the matter-dominated phase ($\omega_{\rm tot} \approx 0$), reaches the threshold for cosmic acceleration at $  -\frac{1}{3}$, and stabilizes around $ -0.65$ at late time and continue with this value to distant future. This interacting model successfully demonstrates a late-time, non-phantom dark energy scenario driven by the curvature sector, avoiding any instabilities associated with phantom behavior. The deceleration parameter  further confirms this, showing transitions consistent with the total EoS evolution, moving from deceleration in the radiation and matter-dominated eras to acceleration in the dark energy era.

\subsection{Comparison of  coincidence \& cosmographic parameters evolution in two models}

\begin{figure}[H]
\centerline{$\begin{array}{cc}
\includegraphics[width=0.4\textwidth]{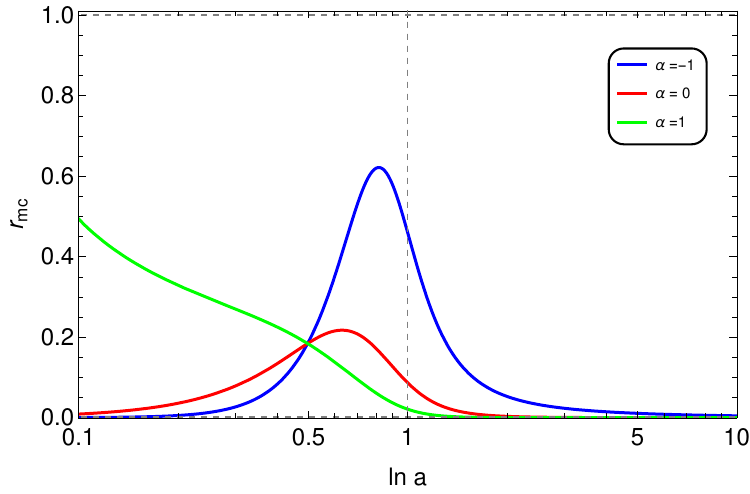}\quad \quad &
\includegraphics[width=0.4\textwidth]{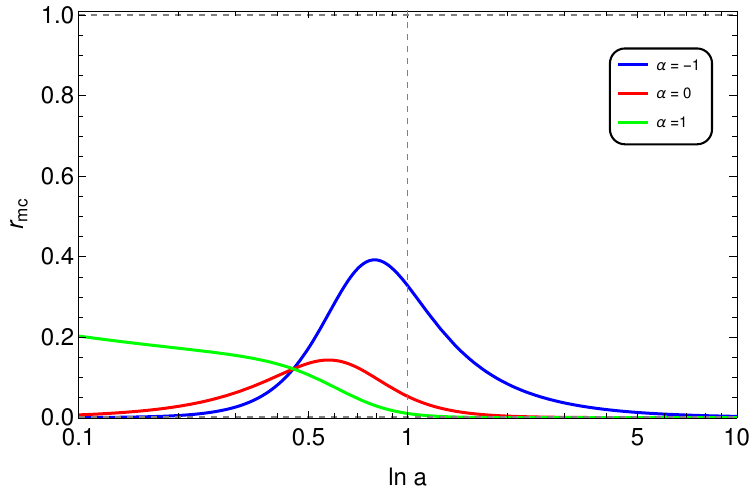}
\end{array}$}

\caption{Evolution plot of the parameter $r_{mc}$ for different values of coupling strength $\alpha$. Left Panel: Generalised $\Lambda$CDM model, Right panel: Power-law model} 
\label{fig:3}
\end{figure}

In fig. \ref{fig:3}, we present a comparative analysis of the dynamical evolution of the matter-to-curvature energy density ratio parameter, $r_{\rm mc}$. The left panel depicts the evolution of the generalized $\Lambda$CDM model, while the right panel corresponds to the power-law model. The parameter $r_{\rm mc}$, defined as  $r_{\rm mc} = \frac{\Omega_{\rm m}}{\Omega_{\rm curv}}$,  quantifies the relative dominance of matter over curvature energy density in the universe. Specifically, $r_{\rm mc} = 0$ corresponds to complete curvature dominance, whereas $r_{\rm mc} \approx 1$ represents a near-equilibrium between matter and curvature energy densities, providing quantitative insights into the dominance of curvature over matter at different stages of cosmic evolution in the respective scenarios. Effect of curvature-matter coupling strength is evident from the $r_{\rm mc}$ profiles for different interaction parameter $\alpha$ values. According to eqs. \eqref{eq:b15} and \eqref{eq:b16}, negative $\alpha$ indicates energy transfer from matter to curvature, while positive $\alpha$ reflects the reverse. Peaks in $r_{\rm mc}$ profiles correspond to epochs of maximum matter density. For $\alpha \sim -1$ or more negative, peak heights approaching towards unity indicate near-equal matter and curvature energy densities. Strongly negative $\alpha$ values lead to late-time phases where matter and curvature remain comparable, though $r_{\rm mc} < 1$ implies curvature dominance. During earlier decelerated phases, low $r_{\rm mc}$ reflects curvature's dominance over matter. For $\alpha = 0$, representing standard $f(R)$ gravity, the $r_{\rm mc}$ profile is symmetric around the epoch of peak matter density. Positive $\alpha$ results in an early non-zero $r_{\rm mc}$, which decreases before rising to a maximum during the epoch of peak matter contribution. The plot also highlights that interactions significantly influence matter and curvature densities at both early and late times, unlike in the absence of interaction.

  The parametric evolution of cosmographic quantities is illustrated in the $j$-$q$ plane for two interacting $f(R)$ gravity models, along with their corresponding non-interacting counterparts that remain consistent with theoretical and current observational bounds on model parameters. In the absence of matter-curvature coupling, viable $f(R)$ models are primarily constrained by cosmological (CMB, BAO, SNe Ia, RSD, WL) and solar System observations through the PPN formalism. For the generalized $\Lambda$CDM model, defined by $f(R) = (R^b - \Lambda)^c$, observational consistency requires parameter ranges such that $b \approx 1$, $c \lesssim 1$, and $bc \approx 1$ (though not exactly equal to unity), with $\Lambda \sim \mathcal{O}(H_0^2)$, while also satisfying the chameleon screening condition $f_{RR} \ll 1/R$ in high-density environments~\cite{KhouryWeltman2004, Brax2008}. In the case of the power-law model, $f(R) = R - \gamma R^n$ \cite{Amendola:2006kh, Starobinsky2007}, viable constraints are found within the range $0 < n < 1$, typically around $n \sim 0.9$~\cite{Carvalho:2008am}, with $\gamma \lesssim 10^{-6}$ (in units of $H_0^{2(1-n)}$) and a suppressed deviation $\gamma R^{n-1} \ll 1$, as supported by Planck and large-scale structure data~\cite{Planck2018, Capozziello2015, Motohashi2013, Dossett2014}. In the left panel, the dynamics of the generalized $\Lambda$CDM model are displayed both with and without matter-curvature coupling. Trajectories deviate from the standard $\Lambda$CDM expectation ($j = 1$), depending on the interaction strength $\alpha$ and the product $bc$. For instance, the green ($\alpha = 1, bc = 1.75$) and blue ($\alpha = 2, bc = -3$) curves illustrate extended dynamical trajectories that cross $j = 1$ and depart from the de Sitter attractor. The red curve corresponds to the non-interacting case favored by observational data. Arrows on the trajectories indicate the evolution from earlier to later cosmic times. The right panel presents the power-law model with interaction (blue curve: $\alpha = 1, n = -1$) compared to the observationally viable non-interacting case (red curve: $\alpha = 0, n = 0.9$). This model shows significant departures from standard $\Lambda$CDM behavior, especially near the epoch of transition from decelerated to accelerated expansion. In the late-time regime ($q < 0$), the evolution of the jerk parameter $j$ effectively captures the influence of curvature-driven acceleration. In both plots, the black dashed curve represents the standard $\Lambda$CDM scenario. All trajectories-whether corresponding to interacting or non-interacting cases-evolve toward the point characterized by unit jerk and deceleration parameter $q = -1$. Collectively, these plots  provide a comparative dynamical perspective on the behavior of interacting versus non-interacting $f(R)$ models, evaluated in the context of their consistency with observational data.

\begin{figure}[H]
\centerline{$\begin{array}{cc}
\includegraphics[width=0.45\textwidth]{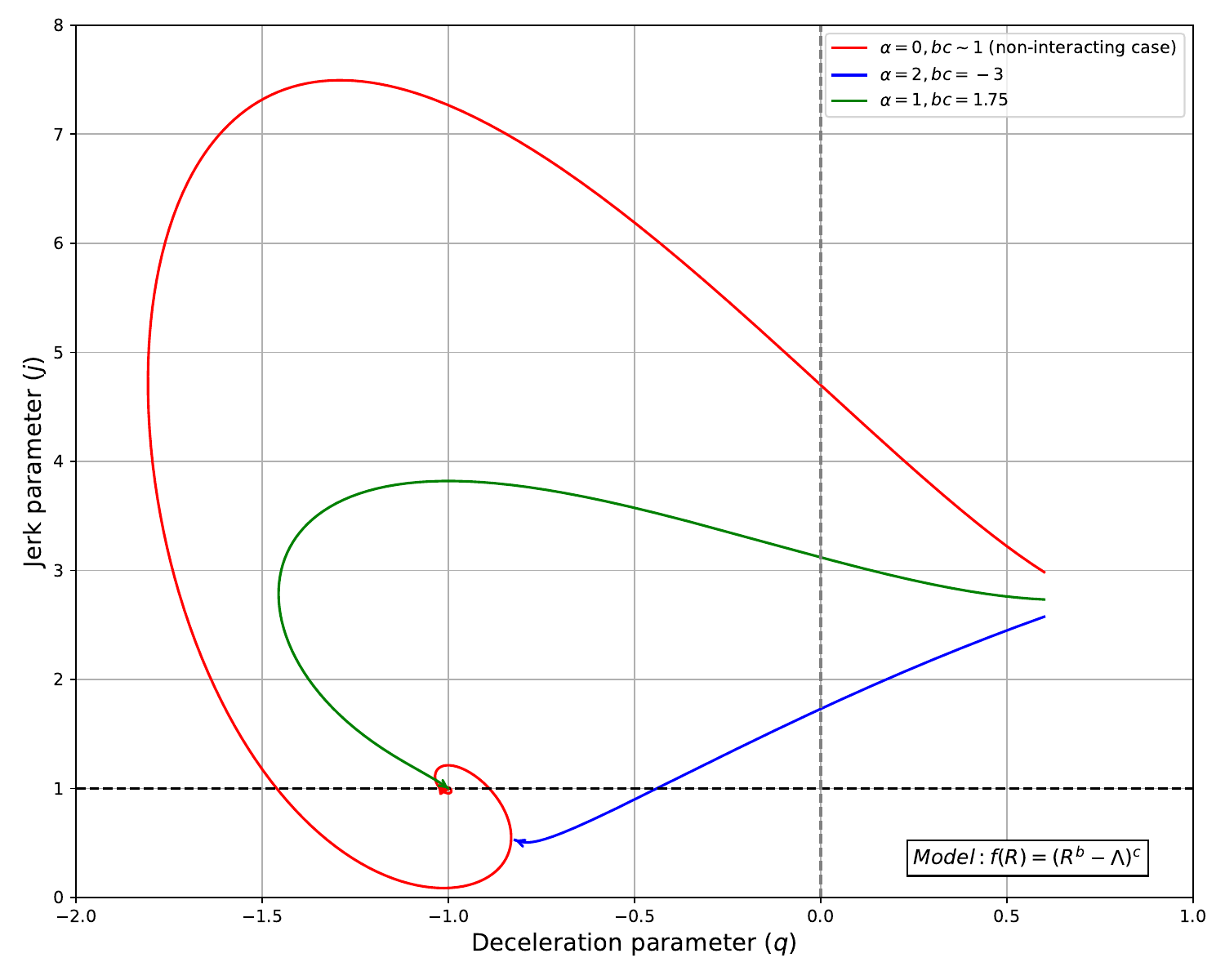}\quad \quad &
\includegraphics[width=0.45\textwidth]{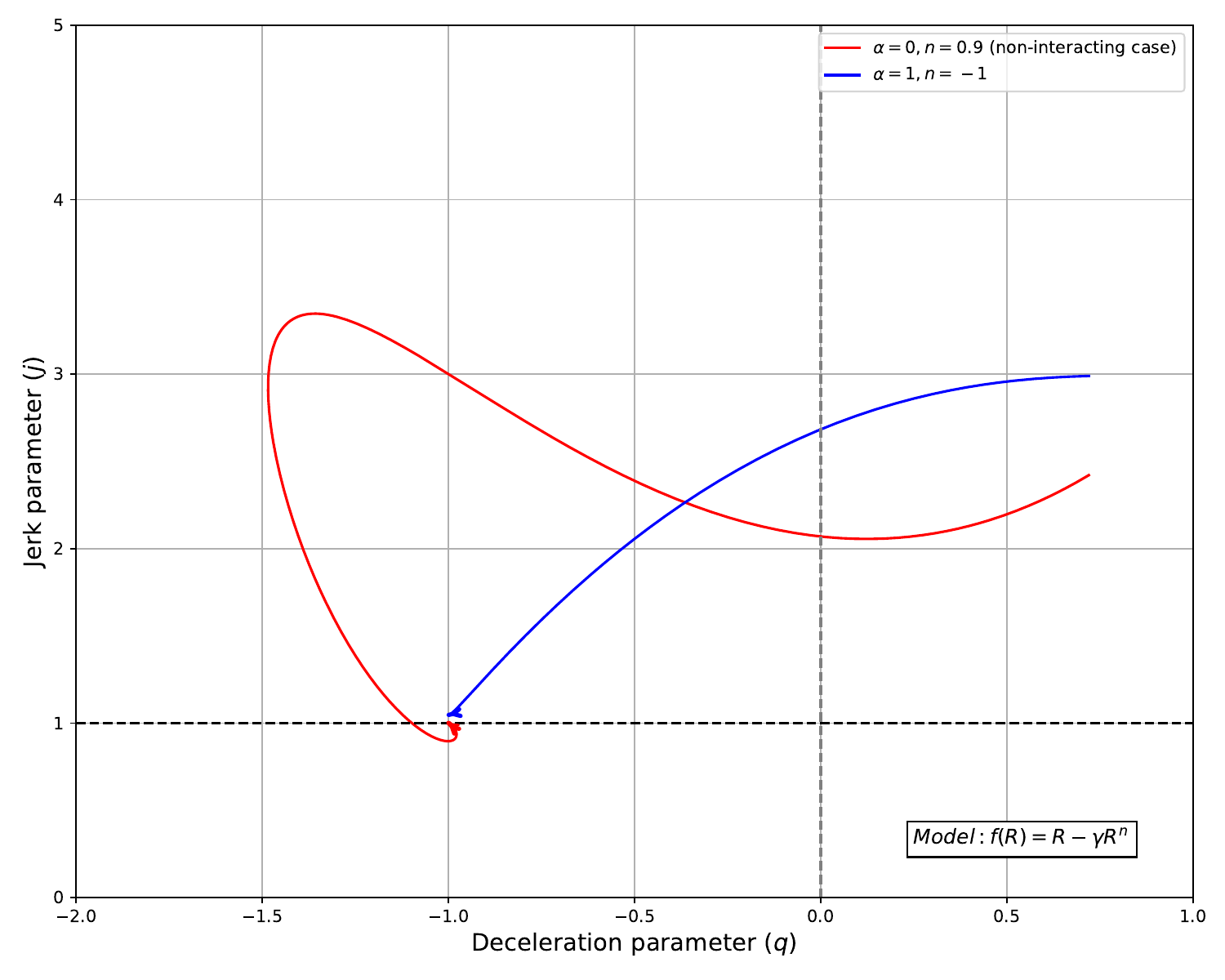}
\end{array}$}

\caption{Parametric plot of jerk parameter with respect to
deceleration parameter. Left Panel: Generalised $\Lambda$CDM model, Right panel: Power-law model} 
\label{fig:4}
\end{figure}

\section{Conclusion}
\label{Sec:4}
In this work, we examine interactions between curvature and matter within viable $f(R)$ gravity models, ensuring their ability to drive cosmic acceleration. At the action level, we assume minimal coupling between matter and curvature, with the radiation component decoupled, preserving its energy-momentum tensor. Interactions are introduced via a source term $\mathcal{Q}$ in the continuity equations for matter and curvature sectors, with opposite signs, ensuring energy conservation in the combined sector. A specific form of $\mathcal{Q}$ is adopted, which compiles both the effect of curvature and matter sectors, modulated by a dimensionless parameter $\alpha$ that controls the interaction strength. The sign of $\alpha$ determines the direction of energy transfer between these two sectors. This particular form of chosen source term can exhibit distinct scaling across cosmic epochs, effectively modeling the transition from radiation to dark energy. In the early universe, it minimizes interaction with radiation while enabling energy exchange between matter and curvature. During the matter-dominated era, it facilitates a gradual transfer of energy from matter to curvature. As the universe enters the dark energy-dominated phase, the term ensures curvature dominance and drives cosmic acceleration. This dynamic behavior allows the coupling term to adapt seamlessly across different epochs, providing a unified framework for describing cosmic evolution. \\

This study establishes a 4-D dynamical framework to describe the curvature-matter interaction scenario. The inclusion of the interaction term $\mathcal{Q}$ modifies the autonomous equations governing the dynamical system, distinguishing it from $f(R)$ theories without such interactions. Critical points of the system were identified and their stability were analyzed for two viable $f(R)$ models: the generalized $\Lambda$CDM model and the power-law model. The interaction parameter $\alpha$ plays a pivotal role, influencing the stability of fixed points and shaping the universe's evolution during various phases. Depending on the choice of the interaction term and considering that our study focuses primarily on the interaction between curvature-driven dark energy and matter, the radiation component has been neglected. Since one of the dynamical variables represents the energy density of radiation, excluding its effect reduces the dimensionality of the phase space from four to three, enabling a depiction of all critical points within this newly constructed 3D phase portrait.\\

Incorporating curvature-matter coupling ($\alpha \neq 0$) in $f(R)$ gravity alters the number of critical points. Depending on the chosen benchmark values, new stable attractors emerge in both models and the nature of the stable de Sitter attractor shifts, allowing for additional stable fixed points. In both types of $f(R)$ gravity models, stable de Sitter attractors emerge: $P_2$ and $P_7$ in the generalized $\Lambda$CDM case, and $P_2$ and $P_5$ in the power-law model, compared to only $P_2$ in the non-interacting case ($\alpha = 0$). A comprehensive stability analysis across various $\alpha$ values and specific model parameters ($bc$ or $n$) reveals intriguing patterns within the stability landscape of the matter-curvature interaction scenario. In both cases, we analyzed the parameter space of the models and coupling parameters at specific critical points, identifying regions where stable and accelerating solutions coexist for certain benchmark values. For the region $1.65 < bc < 2$ and $\alpha > -1.5$, the fixed point $P_2$ exhibits a stable de Sitter-type accelerating solution. For $P_7$, the region is different, as shown in fig. \ref{fig:1a}. The figure clearly illustrates that both critical points can produce stability and acceleration, but at different benchmarks. By selecting two distinct benchmarks within this allowable region, we plotted phase space trajectories, which reveal the movement of the phase flow near the critical points. We identified $P_2$ and $P_7$ as fixed points for their respective benchmarks, while both points exhibit saddle-type behavior at other values. Similarly, in the second case, we identified two fixed points, $P_2$ and $P_5$, and marked their stability and accelerating regions, as shown in fig. \ref{fig:2a}. In the region, $1.3 < n < 2$ and $\alpha > -1.4$, $P_2$ exhibits a stable de Sitter-type accelerating solution, while $P_5$ spans a different region, depicted in fig. \ref{fig:2a}. Since $P_2$ shows behavior similar to the previous model in the chosen benchmark from the allowable region, we shifted our focus to investigate the phase space where $P_5$ shows stable and accelerating behavior. Interestingly, at these benchmark values, $P_2$ becomes a saddle point, with all trajectories in the phase portrait repelling from all other points and moving toward $P_5$, making it the attractor point. \\

 Evolution of cosmological and cosmographic parameters was studied for both models to assess the impact of curvature-driven dark energy on cosmic evolution within the matter-curvature interaction framework. Key parameters, including modified energy densities, the equation of state (EoS) parameter, the matter-to-curvature density ratio ($r_{\rm mc}$),  deceleration ($q$) and jerk ($j$), provide insights into possible evolutionary scenarios. In the first model, at the benchmark values $\alpha = 1$ and $bc = 1.75$, the curvature and matter energy densities initially have equal values. Over time, the curvature energy density increases relative to the matter density. Once the effective equation of state (EOS) parameter crosses $-\frac{1}{3}$, the matter density diminishes to zero. The EOS parameter further crosses $-1$, entering  an unstable phantom region before rebounding and stabilizing at $-1$. The deceleration parameter starts at $+1$ and eventually stabilizes at $-1$, indicating a stable de Sitter accelerating fixed point, $P_2$. At a different benchmark value ($\alpha = 2$ and $bc = -3$), where $P_7$ is stable, the matter density initially dominates over the curvature energy density. Over time, the curvature energy density grows and surpasses the matter density, which eventually tends to zero. The EOS parameter begins at $+\frac{1}{3}$ and stabilizes near $-0.8$, while the deceleration parameter transitions from $+1$ to approximately $-0.6$. For the power-law model, at the benchmark values $n = -1$ and $\alpha = 1$, the stable fixed point $P_5$ is observed. In the early universe, matter density dominates over curvature density but decreases over time, leading to late-time dominance of curvature-driven dark energy. The effective equation of state (EOS) parameter transitions from a radiation-dominated phase to a dark energy-dominated phase, eventually stabilizing near $-0.65$. Across all evolutionary plots, the dynamical behavior of the EOS parameter consistently exhibits non-phantom dark energy. Coincidence ($r_{mc}$) parameter quantifies the dominance of matter over curvature component. Without interactions, $r_{mc}$ exhibits a symmetric profile around the epoch of maximum matter energy density, approaching zero at both early and late times. For $\alpha \neq 0$, positive $\alpha$ enhances matter contributions at earlier epochs, while negative $\alpha$ amplifies matter dominance at later times. Notably, a negative $\alpha$ provides insights into the cosmological coincidence problem in the late universe, which we can see in both models. Parametric evolution of the jerk ($j$) and deceleration ($q$) parameters in interacting generalized $\Lambda$CDM and power-law $f(R)$ gravity models show notable deviations from the standard $\Lambda$CDM. These deviations, driven by modified gravity and energy exchange between dark sectors, lead to a dynamic $j$ instead of the fixed $j = 1$ in $\Lambda$CDM. The resulting trajectories suggest richer late-time dynamics, including possible departures from the de Sitter asymptote and varying effective equations of state, offering an alternative explanation for cosmic acceleration to be tested against data.   Within the framework of the specific interacting curvature-matter scenario considered in this work, deriving stringent observational constraints on the model and coupling parameters necessitates a detailed and systematic analysis of a wide range of cosmological datasets—an endeavor that lies beyond the scope of the present study. Instead, we adopt a dynamical systems approach to constrain the parameter space $(bc, \alpha, n)$ for the two $f(R)$ gravity models under consideration, focusing on the stability of critical points and the realization of a stable late-time accelerated expansion. In addition, we perform a comparative analysis between the interacting models developed in this work and the corresponding non-interacting scenarios, where model parameters are chosen to satisfy both theoretical viability conditions of $f(R)$ gravity and observational consistency. This comparison is presented in the context of parametric evolution plots in the cosmographic parameter space. The results indicate that such interacting curvature-matter $f(R)$ models are capable of producing a stable late-time non-phnatom acceleration while potentially addressing the cosmic coincidence problem. A more detailed investigation into the observational implications of the interaction, along with a comprehensive constraint analysis of the full parameter space, will be carried out in future work.\\

While numerous $f(R)$ models address dark energy, our findings provide qualitative insights and supplementary constraints for evaluating their viability. Each model uniquely influences cosmological perturbations, particularly in structure formation, which is sensitive to dark energy-dark matter interactions in $f(R)$ gravity. These interactions may impact the matter-radiation equality epoch and contribute to anisotropies observable in structure growth. Early universe interactions could also help explain the Hubble tension by reconciling local and CMB measurements. Studying matter perturbations and local gravity approximations in interacting $f(R)$ models offers valuable perspectives on inconsistencies with non-interacting models, while alternative interacting models may reveal further cosmological implications.\\

In conclusion, this study deepens our understanding of modified $f(R)$ gravity, with a focus on curvature-matter interactions. The discovery of stable de-Sitter attractors, intricate stability characteristics, and unique evolutionary trajectories highlights the complexity of the system's dynamics. These results provide valuable insights into cosmological behavior, offering a foundation for exploring the fundamental processes that govern the universe.

 \vspace{1cm}
\paragraph*{Acknowledgement}
The authors are thankful to the referee for the valuable suggestions, which have helped improve the quality and clarity of the manuscript. This work is supported in part by the National Key Research and Development Program of China under Grant No. 2020YFC2201504. 
 
 \vspace{1cm}
\paragraph*{Data Availability Statement }  No data is associated with the manuscript.

\appendix 

\section{A details on weak-field solar system constraints on $f(R)$ gravity models}
\label{Sec:Apna}

We begin by considering the action for $f(R)$ gravity in the Jordan frame under minimal coupling between the metric and matter fields:
\begin{equation}
S = \frac{1}{2\kappa^2}\int d^4x\sqrt{-g}f(R) + \int d^4x~L_M(g_{\mu \nu},\phi _M)\
\end{equation}
where $f(R)$ is a general function of the Ricci scalar and $S_m$ represents the matter action minimally coupled to the metric $g_{\mu\nu}$.

Varying this action with respect to the metric yields the modified field equations:
\begin{equation}
F(R) R_{\mu\nu}-\frac{1}{2}g_{\mu \nu}f(R) + g_{\mu \nu}\square F(R) - \nabla_{\mu}\nabla_{\nu} F(R)  = \kappa^2 \tilde{T}_{\mu\nu}^{(\rm M)}\,
\end{equation}
In this context, $F(R) \equiv \frac{df}{dR}$, and $\tilde{T}_{\mu \nu}^{(\rm M)}$ represents the stress-energy tensor for the radiation and matter components and $\tilde{T}_{\mu \nu}^{(\rm M)} = F T_{\mu \nu}^{(\rm M)}$.

In many formulations, one can reorganize the geometric terms to define an effective curvature energy-momentum tensor $T^{\text{curv}}_{\mu\nu}$ such that:
\begin{equation}
G_{\mu\nu} = \kappa^2 (T^{\text{curv}}_{\mu\nu} + T^{\text{M}}_{\mu\nu}),
\end{equation}
making the gravitational field equations structurally similar to those of general relativity but with an additional curvature-sourced contribution. Here,   $ T^{\text{M}}_{\mu\nu} = T^{\text{m}}_{\mu\nu} + T^{\text{r}}_{\mu\nu}$. Importantly, although $T^{\text{curv}}_{\mu\nu}$ and $T^{\text{m}}_{\mu\nu}$ may not be conserved individually, the total energy-momentum tensor remains conserved:
\begin{equation}
\nabla^\mu (T^{\text{curv}}_{\mu\nu} + T^{\text{m}}_{\mu\nu} + T^{\text{r}}_{\mu\nu}) = 0.
\end{equation}
Where, the radiation part has been individually conserved ($\nabla^\mu T^{\text{r}}_{\mu\nu} = 0 $) in our work.

\subsection*{Local Limit and Justification for Neglecting Curvature-Matter Interaction}

In a cosmological setting, particularly with non-minimal curvature-matter couplings, interactions between $T^{\text{curv}}_{\mu\nu}$ and $T^{\text{m}}_{\mu\nu}$ can dynamically influence the background evolution of the universe. However, for the purposes of Solar System tests and weak-field approximations, the relevant physical context involves a local, static, and spherically symmetric matter source (e.g., a star) embedded within a homogeneous and isotropic cosmological background \cite{Nojiri:2007as,Chiba:2006jp}.

In this regime, we consider perturbations of the Ricci scalar of the form:
\begin{equation}
R(r, t) = R_0(t) + R_1(r), \quad \text{with} \quad R_1 \ll R_0.
\end{equation}
Here, $R_0(t)$ encodes the time-dependent cosmological background, while $R_1(r)$ is the small spatial perturbation due to the local source.

Because the local perturbation is treated as static and spatially dependent, and since the cosmological background (including any curvature-matter interactions) varies only with time, their mutual interaction becomes negligible in the weak-field regime. Consequently, we neglect any cross-interaction terms between the matter and curvature sectors. The trace of the energy-momentum tensor used in the field equations is then effectively split as:
\begin{equation}
T = T_{\text{cos}}(t) + T_s(r), \quad \text{where } T_s \equiv -\rho(r).
\end{equation}

The perturbative analysis, including the derivation of the scalar field's effective mass and the PPN parameter $\gamma$, is thereby governed solely by the local matter source. The background curvature-matter coupling influences only the large-scale cosmological behavior and does not modify the local weak-field equations relevant for Solar System constraints.

\subsection*{Weak-Field Limit of  $f(R)$  Gravity: Theoretical and Mathematical Framework}

\subsection*{Field Equations and Trace Equation}
We consider the modified gravity action:
\begin{equation}
S = \frac{1}{2\kappa^2}\int d^4x\sqrt{-g}f(R) + \int d^4x~L_M(g_{\mu \nu},\phi _M)\
\end{equation}
where $f(R)$ is a general function of the Ricci scalar $R$ and $S_m$ is the matter action. Varying with respect to $g_{\mu\nu}$ yields the field equations:
\begin{equation}
F(R) R_{\mu\nu}-\frac{1}{2}g_{\mu \nu}f(R) + g_{\mu \nu}\square F(R) - \nabla_{\mu}\nabla_{\nu} F(R)  = \kappa^2 \tilde{T}_{\mu\nu}^{(\rm M)}\,
\end{equation}
where $F \equiv df/dR$. Taking the trace of the above equation gives:
\begin{equation}
F R - 2f + 3 \Box F = \kappa^2 T.
\end{equation}

\subsection*{Perturbation Around a Cosmological Background}
We consider a background scalar curvature $R_0(t)$ with cosmological sources $T_{\text{cos}}(t)$, and perturb it by a local, static, spherically symmetric source (e.g., a star) with spatial dependence:
\begin{equation}
R(r, t) = R_0(t) + R_1(r), \quad R_1 \ll R_0.
\end{equation}
The trace equation splits into background and perturbation parts. Subtracting the background part, we obtain the linearized trace equation:
\begin{equation}
3 f_{RR}^0 \Box R_1 - (F^0 - f_{RR}^0 R_0) R_1 = \kappa^2 T_s,
\end{equation}
which becomes, in static limit ($\Box \to \nabla^2$):
\begin{equation}
\nabla^2 R_1 - m^2 R_1 = -\frac{\kappa^2}{3 f_{RR}^0} \rho(r),
\end{equation}
with the effective scalar mass defined by:
\begin{equation}
m^2 \equiv \frac{1}{3 f_{RR}^0} (F^0 - f_{RR}^0 R_0).
\end{equation}

\subsection*{Solution Using Green's Function}
The equation for $R_1$ is of Helmholtz type. The solution, for a localized mass $M$, is:
\begin{equation}
R_1(r) = -\frac{\kappa^2 M}{12\pi f_{RR}^0} \frac{e^{-mr}}{r}.
\end{equation}
In the limit $mr \ll 1$, this reduces to:
\begin{equation}
R_1(r) \approx -\frac{\kappa^2 M}{12\pi f_{RR}^0} \frac{1}{r}.
\end{equation}

\subsection*{Metric Potentials and Modified Gravity Effects}
We work in the Newtonian gauge with the metric:
\begin{equation}
ds^2 = -(1 + 2\phi(r)) dt^2 + (1 - 2\psi(r)) dr^2 + r^2 d\Omega^2.
\end{equation}
Using the linearized field equations:
\begin{align}
F^0 \nabla^2 \phi &= \frac{\kappa^2}{2} \rho(r) + \frac{1}{4} R_1(r), \\
F^0 \nabla^2 \psi &= \frac{\kappa^2}{2} \rho(r) - \frac{1}{4} R_1(r),
\end{align}
we obtain the gravitational potentials outside the source:
\begin{align}
\phi(r) &= -\frac{G_{\text{eff}} M}{r}, \quad G_{\text{eff}} = \frac{\kappa^2}{6\pi F^0}, \\
\psi(r) &= -\frac{G_{\text{eff}} M}{2r}.
\end{align}
This leads to the post-Newtonian parameter:
\begin{equation}
\gamma = \frac{\psi}{\phi} = \frac{1}{2},
\end{equation}
which deviates from GR ($\gamma = 1$) and violates Solar System constraints unless the scalar field is heavy ($m^2 r^2 \gg 1$).

\subsection*{Constraints on Specific $f(R)$ Models}

\paragraph{Model I: $f(R) = (R^b - \Lambda)^c$}

The effective scalar mass squared is:
\begin{equation}
m^2 = \frac{1}{3} \left( \frac{f_R}{f_{RR}} - R \right),
\end{equation}
with:
\begin{align*}
f_R &= cb R^{b-1} (R^b - \Lambda)^{c-1}, \\
f_{RR} &= cb(b-1) R^{b-2} (R^b - \Lambda)^{c-1} + c(c-1)b^2 R^{2b-2} (R^b - \Lambda)^{c-2}.
\end{align*}

\textbf{Constraint:} To ensure $m^2 r^2 \gg 1$ for Solar System scale $r \sim 1\, \text{AU}$, we require:
\begin{equation}
\left| \frac{f_R}{f_{RR}} - R \right| \gg \frac{3}{r^2}.
\end{equation}

Since $R \sim R_0 \sim H_0^2$ and $r \sim \text{AU}$, this implies:
\begin{equation}
\left| \frac{f_R}{f_{RR}} \right| \gg R + \frac{3}{r^2} \sim \frac{3}{r^2}.
\end{equation}

This condition is satisfied if:
\begin{itemize}
    \item $b \approx 1$, $c \lesssim 1$,
    \item $bc \approx 1$,
    \item and $\Lambda \sim \mathcal{O}(H_0^2)$.
\end{itemize}

These constraints ensure that the scalar field is heavy enough and the model reduces to GR at small scales.

\paragraph{Model II: $f(R) = R - \gamma R^n$}

We have:
\begin{align*}
f_R &= 1 - \gamma n R^{n-1}, \\
f_{RR} &= -\gamma n(n-1) R^{n-2}, \\
m^2 &= \frac{1}{3} \left( -\frac{1 - \gamma n R^{n-1}}{\gamma n(n-1) R^{n-2}} - R \right).
\end{align*}

\textbf{Constraint:} Again, require:
\begin{equation}
m^2 r^2 \gg 1 \quad \Rightarrow \quad \left| -\frac{1 - \gamma n R^{n-1}}{\gamma n(n-1) R^{n-2}} - R \right| \gg \frac{3}{r^2}.
\end{equation}

This is satisfied when:
\begin{itemize}
    \item $0 < n < 1$ (typically $n \sim 0.9$),
    \item $\gamma \lesssim 10^{-6}$ in units of $H_0^{2(1-n)}$,
    \item and $\gamma R^{n-1} \ll 1$.
\end{itemize}

These ensure that $f_R \approx 1$, $f_{RR}$ is small, and the scalar field becomes heavy, effectively decoupling from Solar System dynamics.

\section{Conformal Transformation and Scalar-Tensor Equivalence}
\label{Sec:Apnb}
The $f(R)$ gravity theory can be equivalently reformulated as a scalar-tensor theory through a series of well-defined mathematical steps. This transformation highlights the presence of an extra scalar degree of freedom inherent in $f(R)$ models \cite{Capozziello:2006dj}.

\subsection*{ Jordan Frame Action}

The original action for $f(R)$ gravity in the Jordan frame is given by:
\begin{equation}
S = \frac{1}{2\kappa^2}\int d^4x\sqrt{-g}f(R) + \int d^4x~L_M(g_{\mu \nu},\phi _M)\
\end{equation}
where \( f(R) \) is a nonlinear function of the Ricci scalar and \( S_m \) is the matter action minimally coupled to the metric (as chosen in our work).

\subsection*{ Introducing an Auxiliary Field}

To linearize the higher-derivative nature of the gravitational sector, we introduce an auxiliary field \( \chi \) such that:
\begin{equation}
S = \frac{1}{2\kappa^2} \int d^4x \sqrt{-g} \left[ f(\chi) + f'(\chi)(R - \chi) \right] + S_m.
\end{equation}
The equation of motion for \( \chi \) yields \( \chi = R \), restoring the original theory on-shell.

Define:
\[
F(\chi) \equiv f'(\chi), \qquad V(\chi) \equiv \chi f'(\chi) - f(\chi),
\]
to rewrite the action in a Brans–Dicke-like scalar-tensor form.

\subsection*{ Scalar Field Redefinition (Jordan Frame)}

Define the scalar field \( \phi = f'(\chi) \), and express the action as:
\begin{equation}
S = \frac{1}{2\kappa^2} \int d^4x \sqrt{-g} \left[ \phi R - V(\phi) \right] + S_m[g_{\mu\nu}, \Psi_m],
\end{equation}
which corresponds to a scalar-tensor theory in the Jordan frame with Brans–Dicke parameter \( \omega = 0 \).

\subsection*{ Conformal Transformation to Einstein Frame}

We perform a conformal transformation to the Einstein frame via:
\[
\tilde{g}_{\mu\nu} = \phi \, g_{\mu\nu}.
\]
Under this transformation, the Ricci scalar transforms as:
\[
R = \phi^{-1} \left( \tilde{R} - \frac{3}{2\phi^2} (\tilde{\nabla} \phi)^2 + 3 \tilde{\Box} \ln \phi \right),
\]
and the metric determinant transforms as:
\[
\sqrt{-g} = \phi^{-2} \sqrt{-\tilde{g}}.
\]

Substituting into the action yields:
\begin{equation}
S = \int d^4x \sqrt{-\tilde{g}} \left[ \frac{1}{2\kappa^2} \tilde{R} - \frac{3}{4\kappa^2 \phi^2} (\tilde{\nabla} \phi)^2 - \frac{V(\phi)}{2\kappa^2 \phi^2} \right] + S_m[\phi^{-1} \tilde{g}_{\mu\nu}, \Psi_m].
\end{equation}

\subsection*{ Canonical Normalization of the Scalar Field}

Define a canonically normalized scalar field:
\[
\varphi = \sqrt{\frac{3}{2\kappa^2}} \ln \phi, \qquad \text{so that} \qquad \phi = e^{\sqrt{\frac{2\kappa^2}{3}} \varphi}.
\]
Then the kinetic term becomes canonical:
\[
\frac{1}{2} (\tilde{\nabla} \varphi)^2,
\]
and the potential in terms of \( \varphi \) becomes:
\[
U(\varphi) = \frac{V(\phi)}{2\kappa^2 \phi^2}.
\]

\subsection*{Final Einstein Frame Action}

Thus, the $f(R)$ theory is equivalently described by a scalar field \( \varphi \) minimally coupled to gravity in the Einstein frame:
\begin{equation}
S = \int d^4x \sqrt{-\tilde{g}} \left[ \frac{1}{2\kappa^2} \tilde{R} - \frac{1}{2} (\tilde{\nabla} \varphi)^2 - U(\varphi) \right] + S_m[e^{-2\sqrt{\frac{\kappa^2}{6}} \varphi} \tilde{g}_{\mu\nu}, \Psi_m].
\end{equation}

This transformation makes the scalar degree of freedom explicit and simplifies the study of dynamical and observational properties of $f(R)$ theories in a manner analogous to quintessence and Brans–Dicke theories.

\end{document}